\input epsf
\input harvmac
 %
\catcode`@=11
\def\rlx{\relax\leavevmode}                  
 %
 %
 %
\font\tenmib=cmmib10
\font\sevenmib=cmmib10 at 7pt 
\font\fivemib=cmmib10 at 5pt  
\font\tenbsy=cmbsy10
\font\sevenbsy=cmbsy10 at 7pt 
\font\fivebsy=cmbsy10 at 5pt  
\def\BMfont{\textfont0\tenbf \scriptfont0\sevenbf
                              \scriptscriptfont0\fivebf
            \textfont1\tenmib \scriptfont1\sevenmib
                               \scriptscriptfont1\fivemib
            \textfont2\tenbsy \scriptfont2\sevenbsy
                               \scriptscriptfont2\fivebsy}
\def\BM#1{\rlx\ifmmode\mathchoice
                      {\hbox{$\BMfont#1$}}
                      {\hbox{$\BMfont#1$}}
                      {\hbox{$\scriptstyle\BMfont#1$}}
                      {\hbox{$\scriptscriptstyle\BMfont#1$}}
                 \else{$\BMfont#1$}\fi}
 %
 %
 %
 %
\def\inbar{\vrule height1.5ex width.4pt depth0pt}
\def\sinbar{\vrule height1ex width.35pt depth0pt}
\def\ssinbar{\vrule height.7ex width.3pt depth0pt}
\font\cmss=cmss10
\font\cmsss=cmss10 at 7pt
\def\ZZ{\rlx\leavevmode
             \ifmmode\mathchoice
                    {\hbox{\cmss Z\kern-.4em Z}}
                    {\hbox{\cmss Z\kern-.4em Z}}
                    {\lower.9pt\hbox{\cmsss Z\kern-.36em Z}}
                    {\lower1.2pt\hbox{\cmsss Z\kern-.36em Z}}
               \else{\cmss Z\kern-.4em Z}\fi}
\def\Ik{\rlx{\rm I\kern-.18em k}}  
\def\IC{\rlx\leavevmode
             \ifmmode\mathchoice
                    {\hbox{\kern.33em\inbar\kern-.3em{\rm C}}}
                    {\hbox{\kern.33em\inbar\kern-.3em{\rm C}}}
                    {\hbox{\kern.28em\sinbar\kern-.25em{\sevenrm C}}}
                    {\hbox{\kern.25em\ssinbar\kern-.22em{\fiverm C}}}
             \else{\hbox{\kern.3em\inbar\kern-.3em{\rm C}}}\fi}
\def\IP{\rlx{\rm I\kern-.18em P}}
\def\IR{\rlx{\rm I\kern-.18em R}}
\def\Ione{\rlx{\rm 1\kern-2.7pt l}}
 %
 %

 %

\def\intem#1{\par\leavevmode%
              \llap{\hbox to\parindent{\hss{#1}\hfill~}}\ignorespaces}
 %


 %
\newskip\humongous \humongous=0pt plus 1000pt minus 1000pt   
\def\caja{\mathsurround=0pt}
\newif\ifdtup
 %
\def\eqalign#1{\,\vcenter{\openup2\jot \caja
     \ialign{\strut \hfil$\displaystyle{##}$&$
      \displaystyle{{}##}$\hfil\crcr#1\crcr}}\,}
 %
\def\twoeqsalign#1{\,\vcenter{\openup2\jot \caja
     \ialign{\strut \hfil$\displaystyle{##}$&$
      \displaystyle{{}##}$\hfil&\hfill$\displaystyle{##}$&$
       \displaystyle{{}##}$\hfil\crcr#1\crcr}}\,}
 %

 %

 %

 %

 %

 %
 %
 %
 %
\def\,{\hskip1.5pt}           
 %
\let\a=\alpha
\let\b=\beta
\let\c=\chi
\let\d=\delta

\let\g=\gamma

\let\l=\lambda                                   
\let\m=\mu
\let\n=\nu

\let\t=\tau

\let\y=\upsilon                                  

 %
 %
\def\Box{\sqcap\llap{$\sqcup$}}
\def\lapp{\lower.4ex\hbox{\rlap{$\sim$}} \raise.4ex\hbox{$<$}}
\def\gapp{\lower.4ex\hbox{\rlap{$\sim$}} \raise.4ex\hbox{$>$}}
\def\con{\ifmmode\raise.1ex\hbox{\bf*}
          \else\raise.1ex\hbox{\bf*}\fi}
\def\bo{{\raise.15ex\hbox{\large$\Box\kern-.39em$}}}

\def\dual{\relax\leavevmode\lower.9ex\hbox{\titlerms*}}

\let\8=\otimes
 %
 %
 %
 %

\let\2=\underline

 %
\def\dt#1{{\buildrel{\smash{\lower1pt\hbox{.}}}\over{#1}}}

\font\eightrm=cmr8
\def\6(#1){\relax\leavevmode\hbox{\eightrm(}#1\hbox{\eightrm)}}
\def\0#1{\relax\ifmmode\mathaccent"7017{#1}     
                \else\accent23#1\relax\fi}      
\def\7#1#2{{\mathop{\null#2}\limits^{#1}}}      
\def\5#1#2{{\mathop{\null#2}\limits_{#1}}}      
 %

\def\A#1{\left|#1\right|}
 %

 %

 %

 %
\newbox\t@b@x
\def\rightarrowfill{$\m@th \mathord- \mkern-6mu
     \cleaders\hbox{$\mkern-2mu \mathord- \mkern-2mu$}\hfill
      \mkern-6mu \mathord\rightarrow$}
\def\tooo#1{\setbox\t@b@x=\hbox{$\scriptstyle#1$}%
             \mathrel{\mathop{\hbox to\wd\t@b@x{\rightarrowfill}}%
              \limits^{#1}}\,}
\def\leftarrowfill{$\m@th \mathord\leftarrow \mkern-6mu
     \cleaders\hbox{$\mkern-2mu \mathord- \mkern-2mu$}\hfill
      \mkern-6mu \mathord-$}
\def\froo#1{\setbox\t@b@x=\hbox{$\scriptstyle#1$}%
             \mathrel{\mathop{\hbox to\wd\t@b@x{\leftarrowfill}}%
              \limits^{#1}}\,}
 %
\def\frac#1#2{{#1\over#2}}
\def\frc#1#2{\relax\ifmmode{\textstyle{#1\over#2}} 
                    \else$#1\over#2$\fi}           
 %
\def\Claim#1#2#3{\bigskip\begingroup%
                  \xdef #1{\secsym\the\meqno}%
                   \writedef{#1\leftbracket#1}%
                    \global\advance\meqno by1\wrlabeL#1%
                     \noindent{\bf#2}\,#1{}\,:~\sl#3\vskip1mm\endgroup}

\def\QED{\rlx\hfill$\Box$\kern-7pt\raise3pt\hbox{$\surd$}\bigskip}
 %
 %

 %
\def\muthstrut{\vphantom1}
\def\mutrix#1{\null\,\vcenter{\normalbaselines\m@th
        \ialign{\hfil$##$\hfil&&~\hfil$##$\hfill\crcr
            \muthstrut\crcr\noalign{\kern-\baselineskip}
            #1\crcr\muthstrut\crcr\noalign{\kern-\baselineskip}}}\,}

 %
\def\YT#1#2{\vcenter{\hbox{\vbox{\baselineskip0pt\parskip=\medskipamount%
             \def\Box{$\sqcap\llap{$\sqcup$}$\kern-1.2pt}%
              \def\Z{\hfil\vskip-5.8pt}\lineskiplimit0pt\lineskip0pt%
               \setbox0=\hbox{#1}\hsize\wd0\parindent=0pt#2}\,}}}
\def\EU{\rlx\ifmmode \c_{{}_E} \else$\c_{{}_E}$\fi}
\def\TM{\rlx\ifmmode {\cal T_M} \else$\cal T_M$\fi}
\def\TW{\rlx\ifmmode {\cal T_W} \else$\cal T_W$\fi}
\def\CM{\rlx\ifmmode {\cal T\rlap{\bf*}\!\!_M}
             \else$\cal T\rlap{\bf*}\!\!_M$\fi}
\def\hm#1#2{\rlx\ifmmode H^{#1}({\cal M},{#2})
                 \else$H^{#1}({\cal M},{#2})$\fi}
\def\CP#1{\rlx\ifmmode\IP^{#1}\else\IP$^{#1}$\fi}
\def\cP#1{\rlx\ifmmode\IC{\rm P}^{#1}\else$\IC{\rm P}^{#1}$\fi}

\def\sll#1{\rlx\rlap{\,\raise1pt\hbox{/}}{#1}}
\def\Sll#1{\rlx\rlap{\,\kern.6pt\raise1pt\hbox{/}}{#1}\kern-.6pt}
%

 %
 %
\def\ie{\hbox{\it i.e.}}        

\def\CY{Calabi-\kern-.2em Yau}

\def\3{\ifmmode\ldots\else$\ldots$\fi}
\def\Z{\hfil\break\rlx\hbox{}\quad}
\def\3{\ifmmode\ldots\else$\ldots$\fi}
\def\?{d\kern-.3em\raise.64ex\hbox{-}}           
\def\9{\raise.43ex\hbox{-}\kern-.37em D}         

 %
 %

 %

 %

 %
 %
 %
\baselineskip=13.0861pt plus2pt minus1pt
\parskip=\medskipamount
\let\ft=\foot
\noblackbox
\def\SaveTimber{\abovedisplayskip=1.5ex plus.3ex minus.5ex
                \belowdisplayskip=1.5ex plus.3ex minus.5ex
                \abovedisplayshortskip=.2ex plus.2ex minus.4ex
                \belowdisplayshortskip=1.5ex plus.2ex minus.4ex
                \baselineskip=12pt plus1pt minus.5pt
 \parskip=\smallskipamount
 \def\ft##1{\unskip\,\begingroup\footskip9pt plus1pt minus1pt\setbox%
             \strutbox=\hbox{\vrule height6pt depth4.5pt width0pt}%
              \global\advance\ftno by1\footnote{$^{\the\ftno)}$}{##1}%
               \endgroup}
 \def\listrefs{\footatend\vfill\immediate\closeout\rfile%
                \writestoppt\baselineskip=10pt%
                 \centerline{{\bf References}}%
                  \bigskip{\frenchspacing\parindent=20pt\escapechar=` %
                   \rightskip=0pt plus4em\spaceskip=.3333em%
                    \input refs.tmp\vfill\eject}\nonfrenchspacing}}
 %
\def\Afour{\ifx\answ\bigans
            \hsize=16.5truecm\vsize=24.7truecm
             \else
              \hsize=24.7truecm\vsize=16.5truecm
               \fi}
\catcode`@=12
\def\npb#1(#2)#3{{ Nucl. Phys. }{B#1} (#2) #3}
\def\plb#1(#2)#3{{ Phys. Lett. }{#1B} (#2) #3}
\def\pla#1(#2)#3{{ Phys. Lett. }{#1A} (#2) #3}
\def\prl#1(#2)#3{{ Phys. Rev. Lett. }{#1} (#2) #3}
\def\mpla#1(#2)#3{{ Mod. Phys. Lett. }{A#1} (#2) #3}
\def\ijmpa#1(#2)#3{{ Int. J. Mod. Phys. }{A#1} (#2) #3}
\def\cmp#1(#2)#3{{ Commun. Math. Phys. }{#1} (#2) #3}
\def\cqg#1(#2)#3{{ Class. Quantum Grav. }{#1} (#2) #3}
\def\jmp#1(#2)#3{{ J. Math. Phys. }{#1} (#2) #3}
\def\anp#1(#2)#3{{ Ann. Phys. }{#1} (#2) #3}
\def\prd#1(#2)#3{{ Phys. Rev.} {D\bf{#1}} (#2) #3}

\nopagenumbers\abstractfont\hsize=\hstitle
\null
\rightline{\vbox{\baselineskip12pt\hbox{CALT-68-2164}
                                  \hbox{NSF-ITP-98-025}
                                 \hbox{hep-th/9803168}}}%
\vfill

\centerline{\titlefont On the Complementarity of F-theory,}\vskip2pt
\centerline{\titlefont Orientifolds, and Heterotic Strings}
\abstractfont\vfill\pageno=0

\vskip-0.3cm
\centerline{Per Berglund}                                \vskip-.2ex
 \centerline{\it Institute for Theoretical Physics}          \vskip-.4ex
 \centerline{\it University of California}         \vskip-.4ex
 \centerline{\it Santa Barbara, CA 93106, USA}       \vskip-.4ex
\centerline{and}
\centerline{Eric G. Gimon\footnote{$^{}$}
      {Email: berglund@itp.ucsb.edu, egimon@theory.caltech.edu}}  \vskip-.2ex
 \centerline{\it California Institute of Technology}         \vskip-.4ex
 \centerline{\it Pasadena, CA 91125, USA}       \vskip-.4ex
\vfill
\vskip-0.3cm
\vbox{\narrower\baselineskip=12pt\noindent
We study F-theory duals of six dimensional heterotic vacua in extreme
regions of moduli space where the heterotic string is very strongly
coupled.  We demonstrate how to use orientifold limits of these
F-theory duals to regain a perturbative string description.  As an
example, we reproduce the spectrum of a $T^4/\ZZ_{4}$ orientifold
as an F-theory vacuum with a singular $K3$ fibration.  We relate this
vacuum to previously studied heterotic $E_8\times E_8$
compactifications on $K3$.}  

\Date{\vbox{\line{3/98 \hfill}}}

\vfill\eject
\baselineskip=14pt plus 1 pt minus 1 pt

\newsec{Introduction}

\lref\vafa{C. Vafa, {\it Evidence for F-Theory}, hep-th/9602022,
\npb{469} (1996) 403.}
\lref\schwarz{J. H. Schwarz, {\it An $SL(2,\ZZ)$ Multiplet of Type IIB 
Superstrings}, hep-th/9508143, \plb{360} (1995) 13.}
\lref\hulltownsend{C.M. Hull, P.K. Townsend, {\it Unity of Superstring
Dualities}, hep-th/9410167, \npb{438} (1995) 109.}

The interest in duality focuses on the possibility of finding
an alternative description of strong coupling phenomena by means of a
map to a theory which is weakly coupled.  Using this idea all known
perturbative string theories can be related via duality
transformations.  One such theory, the type IIB string theory,
is of particular interest due to a conjectured
non-perturbative $SL(2,\ZZ)$ self-duality~\refs{\hulltownsend,\schwarz}, 
which takes the string coupling $g$ to $1/g$. 	From F-theory, this 
self-duality 
can be used to generate a rich class of non-perturbative vacua, exact up to
corrections of order the string scale. The basic idea is to consider
an artificial twelve dimensional space and compactify it on a
Calabi--Yau manifold which is an elliptic fibration~\vafa. Then, $\t$ of the
torus, with its natural $SL(2,\ZZ)$ action, is 
identified with the complexified axion-dilaton scalar, $\t = a +
ie^{-\phi}$.  In effect, the geometric machinery 
used in constructing Calabi--Yau manifolds generates the analyticity needed for
exact quantum results.  

	There are some operational limits, however, in manipulating F-theory
vacua.  Many vacua of interest are represented via extremely
degenerate geometries.  For these geometries the non-geometric moduli coming 
from the F-theory 7--brane gauge bundles, e.g. Wilson lines, become hard to
identify because the 7--branes themselves lie on very degenerate surfaces (for
example complex hyperboloids which have degenerated to intersecting planes).
Another issue with F-theory vacua is that they receive stringy ${\alpha}'$
corrections.  It would be interesting to get a handle
on these corrections, as this would allow for a more complete picture.  
Fortunately,  F-theory vacua are enmeshed in a web of dualities which can
resolve some of these issues.  Of particular interest is a chain of 
dualities which relates some F-theory vacua to (perturbative)
heterotic vacua.  Sen  
considered this particular chain of dualities in eight dimensions, relating
F-theory on $K3$ to heterotic string theory on $T^2$~\ref\seneightd{A. Sen,
{\it F-theory and Orientifolds}, hep-th/960515,\npb{475} (1996) 562
.}.~\foot{In his original paper, Vafa conjectured a duality between
F-theory on an elliptically fibered $K3$ and the heterotic theory on
$T^2$~\vafa.}   He demonstrated how one could move to a region of
parameter space where the base of the $K3$ resembled a IIB
orientifold.  He used T-duality to relate this 
orientifold to Type~I on $T^2$, then applied Heterotic--Type~I 
duality~\ref\edjoe{J.Polchinski, E. Witten, {\it Evidence For Heterotic--Type~I
Duality}, hep-th/9510169, \npb{460} (1995) 525-540 .}.  Each link in this 
chain, reliable in its own individual region of parameter space, yields 
non-perturbative information on the other links; and there is sufficient  
overlap for us to trust this information.

\lref\vafamori{D. Morrison, C. Vafa,
{\it Compactification of F-theory On Calabi--Yau Threefolds - I},
hep-th/9602114, \npb{473} (1996) 74.}

\lref\vafamorii{D. Morrison, C. Vafa,
{\it Compactification of F-theory On Calabi--Yau Threefolds - II},
hep-th/9603161, Nucl. Phys. B476 (1996) 437-469.}

\lref\senstuff{A. Sen, {\it F-theory and the Gimon-Polchinski 
Orientifold}, hep-th/9702061, \npb{498} (1997) 135.}

	The chain of dualities connecting F-theory to heterotic vacua 
becomes much richer when we consider six dimensional
compactifications~\refs{\vafamori,\vafamorii}.  
In six dimensions there exists a much broader range of 
perturbative and non-perturbative behavior over which to test the
predictive power of these dualities.  For example, for the 
$Spin(32)/\ZZ_2$ heterotic string on $K3$ the possibility of 
``small instantons''~\ref\edsmall{E. Witten, {\it Small
Instantons in String Theory}, hep-th/9511030, \npb{460} (1995) 541-559 }  
yielding extra non-perturbative gauge groups arises.  These are easier 
understood in terms of D5--branes in the context of dual Type~I 
compactifications such as $\ZZ_2$ orientifolds, the so called
GP-models~\ref\ericjoe{E. G. Gimon, J. Polchinski, 
{\it Consistency Conditions for Orientifolds and D-Manifolds}, hep-th/9601038, 
\prd{54} (1996) 1667 .}.  
In a careful
study~\senstuff, Sen showed how these 
orientifolds are in fact T-dual to limits of F-theory vacua involving an 
elliptic Calabi--Yau, ${\cal M}_1$, with base $\IP^1\times\IP^1$.
Type~I/heterotic S-duality allows us to use this map to relate a 
$Spin(32)/\ZZ_2$ compactification on $K3$ to F-theory on ${\cal M}_1$.
This duality is of particular interest because, as we will demonstrate below,
F-theory treats both the perturbative and non-perturbative gauge enhancements 
of its dual heterotic compactification on an equal footing.

	Let us examine the duality between F-theory on ${\cal M}_1$ and 
the heterotic string on $K3$ in terms of the eight dimensional duality 
mentioned previously.  If we think of ${\cal M}_1$ as a $K3$
surface fibered over $\IP^1$, then this eight dimensional duality
maps F-theory on such a $K3$ fibration to a heterotic compactification
involving a $T^2$ fibered over the same $\IP^1$.  Here the geometry of
the $K3$ fibers encode the perturbative gauge group.  There are
two ways to pick the base $\IP^1$ for ${\cal M}_1$, implying the
existence of two different $K3$ fibrations.  These two $K3$ fibrations lead 
to two different descriptions of the heterotic string on $K3$, with different
perturbative gauge groups.  From the point
of view of F-theory, we can think of each of these strings in terms of a 
D3-brane wrapping one or the other of the $\IP^1$ embedded in ${\cal M}_1$.
Each will become weakly coupled when the $\IP^1$ that it wraps becomes very
small relative to the other $\IP^1$.  In this way we recover the 
S--duality of Duff, Minasian and Witten~\ref\dmw{M.J. Duff, R. Minasian and
E. Witten, {\it Evidence for Heterotic/Heterotic Duality}, hep-th/9601036,
\npb{452} (1996) 261}.  Of course, the F-theory limit is best understood when
both $\IP^1$s are large.  This corresponds to treating both
perturbative and non-perturbative gauge groups on an equal footing in
either of the heterotic dual representations.   

\lref\aspin{P. S. Aspinwall, {\it Aspects of the Hypermultiplet Moduli
Space in String Duality}, hep-th/9802194.}

What happens when both $\IP^1$s are small?  The theory receives 
${\alpha}'$ corrections large enough  that F-theory is no longer valid.
But the heterotic dual is not weakly coupled in any description!
Fortunately, in certain regions of the complex structure moduli space
where the $\t$ 
parameter does not vary too much, we can think of the F-theory
compactification in terms of orientifolds and get a good perturbative
description.  In addition, the orientifold description gives us
a better handle on those hypermultiplet moduli which are not, strictly
speaking, directly encoded in the Calabi--Yau 3-fold description of the
F-theory vacuum (\ie, Wilson lines on the 7--branes)~\foot{For a
recent discussion of the hypermultiplet moduli space in the context of
F-theory, see~\aspin.}. Here, we see
the advertised complementarity between F-theory, orientifolds, and the 
heterotic string come in to play.

\lref\ami{O. J. Ganor, A. Hanany, {\it Small $E_8$ 
Instantons and Tensionless Non-critical Strings}, hep/th9602120,
\npb{474} (1996) 122.}

\lref\sixcomments{N. Seiberg and E. Witten, {\it Comments on
String Dynamics in Six Dimensions}, hep-th/9603003, \npb{471} 
(1996) 121.}

\lref\wittenphase{E. Witten, {\it Phase Transitions In
M-Theory And F-Theory}, hep-th/9603150, \npb{471} (1996) 195.}

	So far, we have only used this complementarity to examine possible
perturbative and non-perturbative gauge groups.  The heterotic theory
exhibits another interesting class of non-perturbative effects when
compactified 
on $K3$.  If we look at $E_8 \times E_8$ heterotic vacua with instanton numbers
\foot{For perturbative heterotic vacua, a total of 24 instantons
is required to compensate for the curvature of $K3$} $(12-n,12+n)$
there exists a strong coupling singularity~\dmw.  
Morrison and Vafa explained~\vafamori\ how these heterotic
singularities correspond, in their F-theory dual  
vacua, to the collapse of an exceptional divisor in the F-theory
base~\vafamori.   We
shall only concern ourselves with such divisors of self-intersection number
$-1$ or $-2$, but for a broader class of F-theory vacua.  

	Using the results of Witten~\wittenphase\ (see also \refs{\vafamori,
\vafamorii}), we can interpret the shrunken divisors as follows.  The
collapse of a divisor with self-intersection $-1$ leads to a phase transition
involving a non-critical tensionless string carrying a rank eight current
algebra. (This string is a D3-brane wrapped around the divisor in question).  
This leads to the possibility of a transition to a ``Higgs'' phase with
one less six dimensional tensor multiplet and 29 extra
hypermultiplets.   The exceptional divisor becomes a generic point on 
the base, with no new singularity.  If we take a generic 
F-theory 3-fold compactification,
we can blow up any point on its base, as long as we preserve the 
Calabi--Yau condition.
This transition was first 
understood by Ganor and Hanany~\ami\ (see also \sixcomments) in terms of 
the heterotic string theory
as a small $E_8 \times E_8$ heterotic instanton which 
shrinks and opens up the possibility, via a phase transition, of a "Coulomb"
branch with an extra ${\cal N}=1$ 
tensor multiplet associated with an M-theory 5-brane.

	The collapse of a divisor with intersection number -2
leads to very different
physics~\refs{\vafamori,\sixcomments,\wittenphase,\vafamorii}. 
   First of all, the divisor is blown down to
an $A_1$ singularity in the base.  There is no possibility of a phase 
transition; this is a true boundary in the moduli space of the relevant six
dimensional theory.  Second, because the local geometry of an $A_1$ singularity
is hyper-K\"ahler, the fibration will be trivial in a neighborhood of the 
collapsing divisor.  This means that locally the theory behaves like IIB
at an $A_1$ singularity.  Thus the D3-brane wrapped on this divisor
yields a tensionless non-critical string with twice the supersymmetry.
It will couple to an ${\cal N}=2$ six dimensional tensor multiplet
with five scalars.  In terms of the low energy six dimensional theory, this 
means that we need to tune both an ${\cal N}=1$ hypermultiplet and an
${\cal N}=1$ tensor multiplet to reach this boundary in moduli space.  In 
the dual heterotic theory, there are two ways to reach this type of boundary in
moduli space. 
As introduced above, we can understand this collapse as a strong coupling 
singularity.  Alternatively, when the heterotic theory has M-theory 5-branes
(the heterotic theory is no longer truly perturbative), the same
type of boundary will be reached when 5-branes come
together~\refs{\ami,\sixcomments}.   Neither of these
scenarios is well understood in a perturbative string expansion, though, as
there are large values of the string coupling involved.  A new weakly coupled
stringy description is necessary.  

	One way to get such a weakly coupled stringy description of the physics
near a boundary in moduli space is to use the complementarity between F-theory
and orientifolds.  By this we mean that we will find an orientifold which
describes a limit of F-theory on an elliptic
Calabi--Yau threefold.  In this limit, exceptional divisors of
self-intersection 
number -2 in the base will shrink to zero size.  This paper will study
the limit  
above for a Calabi--Yau space, ${\cal M}_2$. We will denote the (singular)
manifold, in which
the exceptional divisors with self-intersection number -2 have
collapsed, by $\overline {\cal M}_2$. This latter model
will be shown to be the F-theory vacuum corresponding to a $T^4/\ZZ_4$ 
orientifold of Type~IIB~\ref\clifferic{E. G. 
Gimon, C. V. Johnson, {\it K3 Orientifolds}, hep-th/9604129,\npb{477} (1996) 
715.}~\ref\clifferictwo{E. G. Gimon, C. V. Johnson, {\it Multiple Realizations
of ${\cal N}=1$ 
Vacua in Six Dimensions}, hep-th/9606176, \npb{479} (1996) 285-304 .}.
In section~2, we will review the properties of this orientifold, as well
as the $\ZZ_2$ orientifold.  We will also focus on the T--duality which takes 
the general class of orientifolds of this type from a configuration with  
D9--branes and D5--branes (appropriate for Heterotic--Type~I duality) to one 
with D7--branes (such as are found in F-theory).  This will set the stage for 
section~3, where we will review the analysis of how a GP $\ZZ_2$ orientifold
can be seen as a limit of ${\cal M}_1$~\senstuff.  
We will then show in section~4 how 
this leads to a natural construction of the space 
$\overline {\cal M}_2$, corresponding
to the $T^4/\ZZ_4$ orientifold, and how the moduli spaces match.  

	Having found the proper match between F-theory compactification and 
orientifold, we will make use of the complementarity of these two descriptions.
The boundaries in the ${\cal M}_2$ moduli space involving ${\cal N}=2$ 
tensionless strings will be made evident in the weakly coupled orientifold 
string description.  In section~5 we will return to the notion of using 
alternate $K3$ fibrations of the F-theory threefold, ${\cal M}_2$, to
find weakly coupled stringy 
descriptions of different regions of its moduli space.  In this manner, we will
demonstrate how to recover a new description of 
$\overline {\cal M}_2$, such that its 
duality with the heterotic $E_8\times E_8$ theory on $K3$ with
instanton embedding (10,10)  becomes evident.

\newsec{Description of Orientifolds}

\lref\bsag{M. Bianchi, A. Sagnotti, 
{\it Twist Symmetry and Open String Wilson Lines}, 
Nucl. Phys. B361 (1991) 519-538 .}

\lref\attishtwo{A. Dabholkar, J. Park,
{\it Strings on Orientifolds}, 
hep-th/9604178, \npb{477} (1996) 701-714 .}

\lref\attishone{A. Dabholkar, J. Park,
{\it An Orientifold of IIB Theory on K3},
hep-th/9602030, \npb{472} (1996) 207.}

\lref\sengen{A. Sen, {\it Orientifold Limit of F-theory Vacua},
hep-th/9702165, Phys. Rev. D{\bf 55} (1997) 7345-7349 .}

	We are interested in studying IIB orientifolds, such as those 
described in~\refs{\clifferic,\attishtwo}, 
in terms of F-theory on 
elliptically fibered Calabi--Yau 3--folds.  More specifically, we will extend 
the F-theory analysis~\senstuff\ on the GP $\ZZ_2$ 
orientifolds~\ericjoe\ to the $\ZZ_4^A$ family of orientifolds
of~\clifferic.  These orientifolds are the simplest  
generalization of the GP $\ZZ_2$ orientifolds which exhibit 
new behavior.  In the 7--brane picture, which is most useful for
connecting with the
F-theory formalism, this new behavior is manifested in two ways.  First, these
$\ZZ_4$ orientifolds contain not only the $O7_2$ planes of the $\ZZ_2$ models 
($O7_2$ 
planes are orientifold 7--planes, with a deficit angle $\pi$), but also 
orientifold points (actually orientifold 5-planes) and $O7_4$ planes
($O7_2$ planes  
with further identifications due to the presence of orientifold points on 
their world-volume). At strong 
coupling, the $O7_2$ planes can be resolved into two
7--branes with a coupling  
dependent separation~\seneightd~\foot{For a more general analysis of
$O7_2$ planes in 
F-theory see also~\sengen .}.  We will be interested in how this analysis 
extends to the more general $O7_4$ planes and orientifold points found in 
the $\ZZ_4$ models.
Second, these $\ZZ_4$ models have closed string spectra which contain extra 
chiral tensor multiplets \foot{In fact a subset of the $\ZZ_4^A$
models~\clifferic\  (as well as a subset of the GP models)
were first discovered by Bianchi and Sagnotti~\bsag\ as 
early examples of models with extra chiral tensor multiplets.}.
This complicates their relation to potential dual heterotic theories 
(which perturbatively contain no such extra tensors) and has
been the basis for some interesting predictions~\clifferictwo.  One aim
of this work is to put these predictions on a firmer footing.

\subsec{The Transition to 7--branes}

\lref\TK{P.S. Aspinwall, D.R. Morrison,
{\it String Theory on K3 Surfaces}, hep-th/9404151, 
Mirror Symmetry II (B. Greene and S.-T. Yau, eds.),
International Press, Cambridge, 1997, pp. 703-716. }

\lref\berkooz{M. Berkooz, R. G. Leigh, J. Polchinski, J. H. Schwarz,
N. Seiberg, E. Witten,
{\it Anomalies, Dualities, and Topology of D=6 N=1 Superstring Vacua},
hep-th/9605184, Nucl. Phys. B475 (1996) 115.}

	As mentioned earlier, in order to connect Type I orientifolds with 
D9-branes and D5--branes~\refs{\ericjoe,\clifferic,\attishone,
\attishtwo} to F-theory on Calabi--Yau 3--folds, we will look at 
dual models which contain only D7--branes, natural objects in F-theory.
We do this by T--dualizing along the 67-plane \foot{By T-duality, we mean
the T-duality inherited from the covering space $T^4$ which in the smooth
$K3$ should correspond to T-duality as defined in ref.~\TK .}, which 
leaves us with D7--branes along the $01234567$ directions (call them 
7--branes), and D7-branes along the $01234589$ directions  
(we denote them 7'--branes).  
In addition, there will be orientifold 7--planes parallel to both the 7 
and 7'--branes.  It is useful to consider what happens when we make the 
transition from a 5-9 picture to a 7-7' picture.  Both pictures describe  
identical six dimensional ${\cal N}=1$ theories spanning the $012345$ 
directions. Physically equivalent excitations, however, arise from quite
different sources. 

	The map between 5-9 degrees of freedom and 7-7' degrees of freedom 
is very simple for the open string spectrum.  This map is just inherited from 
the T--duality transformations of the original $T^4$.  That is, the D5--branes
become 7--branes and the D9--branes become 7'--branes.  The coordinates of the 
D5--branes on the original $T^4$, which form complete $D=6, \, {\cal N}=1$ 
hypermultiplets, now split into two scalars each from the 89 coordinate of the 
7--brane, and two scalars each from the Wilson lines of the 7--brane gauge 
theory around the 6 and 7 directions.  There is further enhancement to matrix 
valued scalars when several 7--branes sit atop each other. 

A minor subtlety arises when the orientifold projection of the
underlying  $T^4$ 
includes elements which project out 1-cycles.  Naively this could preclude 
Wilson lines on the 7--branes.  These remain, however, because the same 
element which removes 1-cycles has a non-trivial action on the Chan-Paton 
gauge bundle living on the 7--branes.  As one might expect, the combined 
geometric and gauge action of the orientifold group on the 7--brane Wilson 
lines yields exactly the same projection as we had for the 67 coordinates 
of the original D5--branes.   
Schematically, we can attribute the existence of this continuous, though
disconnected, moduli space of flat connections on the 7--branes to the fact 
that they are pierced by orientifold planes.  This allows for non-trivial 
monodromies of the 7--brane gauge bundle about the puncture points.
Similarly, Wilson lines from the D9--branes become hypermultiplets for 
the 7'--branes, parameterizing their position in the 67 directions and
their Wilson lines along the 89 directions~\foot{One might wonder why there
are Wilson lines for  
the original D9--branes on a $T^4$ orientifold.  Again these arise from the
action of the orientifold on the Chan-Paton gauge bundle.  The schematic 
picture for this moduli space for the gauge bundle is  quite different from 
the 7-7' case.  In the 5-9 picture the presence of abelian instantons at the 
core of orientifold points indicates that we are discussing a moduli space of 
{\it{curved}} connections.  There exist continuous moduli describing how
abelian instantons are embedded relative to each other in the 9-brane gauge
group (see ref.~\berkooz).}.

	The 67 T--duality operation is not nearly so simple for the closed 
string sector of our theory.  For the untwisted sector there is again a legacy
from the original $T^4$.  T--duality mixes the components of the metric and
various forms of the Type IIB theory aligned partly or fully along the compact 
directions.  More subtle is understanding the fate of the twisted
sectors after T--duality, as these are added only after the orientifold 
projections.  There is strong evidence from anomaly analysis 
\foot{See for example Berkooz et. al.~\berkooz .} 
that twisted closed string modes localized at one fixed 
point undergo a discrete Fourier transform~\ref\joeprivate{J. Polchinski, 
private communication.}. They are mapped to a linear combination of 
modes localized at separate fixed points in the 67 plane.  
At this point, one might be tempted to forget about the 5-9 picture
and just derive the complete 7-7' spectrum by applying the appropriate 
orientifold projection to the Type IIB string on $T^4$. However, 
it is much simpler 
to understand the interplay of Higgsing patterns with blow-up modes 
of orbifold points in the 5-9 picture. So we will need to 
keep the relationship between the 5-9 picture and the 7-7' picture in the back
of our mind.

\subsec{The Transformed $GP$ Models}

\lref\clifflast{C. V. Johnson, {\it Anatomy of a Duality},
hep-th/9711082.}

	We will use the GP models~\ericjoe\ to illustrate
the relationship between the 5-9 picture and the 7-7' picture.  Here
we start with IIB string theory on $T^4$.  The 5-9  
picture orientifold group is:
\eqn\omega{
\{1,\Omega, \Omega R_{6789}, R_{6789}\},
}
where $R$ is the reflection along the subscripted axes.  As mentioned 
previously, this represents Type I strings on a $K3$ surface in the 
$T^4/\ZZ_2$ orbifold limit.  The closed string sector 
can be easily computed.  The graviton yields the $D=6$ graviton 
and ten scalars describing the shape and size of the original $T^4$.  It also 
yields three scalars from the singular two--cycles at each of the sixteen
fixed points. 
The Ramond--Ramond two--form yields one $D=6$ anti--self--dual and one
self--dual tensor.  The former joins with the graviton to form the
bosonic part of the gravity 
multiplet while the later fills out a chiral ${\cal N}=1$ tensor multiplet 
with the dilaton.  The two-form also yields six scalars representing fluxes on 
the original $T^4$, and a scalar for each singular two--cycle.  In total, 
the $D=6$ 
theory thus has a gravity multiplet along with a tensor multiplet and 20  
hypermultiplets (the 80 moduli for $K3$).  Because of the  curvature
of $K3$, we
also expect the 9--brane gauge bundle to have 24 instantons.  Eight of these
are realized as independent D5--brane units with $SU(2)$ gauge group,
while the 
other 16 consist of abelian instantons located at the core of each $\ZZ_2$ 
fixed point~\refs{\ericjoe,\berkooz} (see also ref.~\clifflast).    

	Now we T--dualize along the 67 directions to get a 7-7' orientifold.
The new orientifold group is
\eqn\omegadual{
\{1,\Omega (-1)^{F_L} R_{67}, \Omega (-1)^{F_L} R_{89}, R_{6789}\}, 
}
where $F_L$ is the left moving fermion number.  Calculating the closed string
spectrum is now slightly more complicated. The orientifold
group in~\omegadual\ can no longer be factorized into a subgroup
acting exclusively on the worldsheet times a subgroup acting on the
target space as in~\omega. The $R_{6789}$ element 
still gives us a $T^4/\ZZ_2$ orbifold limit of $K3$ as the underlying geometry.
But now the $(-1)^{F_L} R_{67}$ element will freeze four of the
$T^4$ metric  
moduli plus one modulus for each singular two--cycle, allowing us to factor 
$T^4$ as $T^2 \times T^2$.  The NS--NS and R--R two--forms are odd under  
$\Omega (-1)^{F_L}$ but since four of the six two--cycles from $T^4$
and all sixteen of the 
singular two--cycles are odd under $R_{67}$, each of the two--forms
will still yield 4 + 16 scalars.  Finally, the R--R zero--form and
self--dual four--form each contribute one scalar.  

	Summarizing, in the untwisted closed string sector we
get three hypermultiplets, each of which has two scalars from the $K3$
metric and two more 
scalars from the R--R and NS--NS two--forms.  There is also one
hypermultiplet with a scalar from the NS--NS two-form and one from
each of the R--R zero, two, and four--form respectively.
This last hypermultiplet is the ``universal'' hypermultiplet which contains 
the volume of the $K3$ (coming from the R--R four--form).  Each of the sixteen 
twisted sectors will contribute one hypermultiplet with two geometric scalars 
from the metric, and one scalar each from the R--R and NS--NS two-forms, 
so-called ``theta angles''.  Note that this means that the fixed points
of $T^4/\ZZ_2$ can not be completely resolved.  

	We now see how the hypermultiplet moduli space of this orientifold
can be matched to the complex deformations of F-theory on a particular 
Calabi--Yau 3--fold, ${\cal M}_1$.  Except for the ``universal''
hypermultiplet,  a special case in F-theory (and M-theory), 
every hypermultiplet can be split into a pair of complex scalars. One
complex scalar comes from either geometric deformations of the
orientifold or from a 7--brane position, and the other scalar from
two--form fluxes or 7--brane Wilson lines.   
The first half of each hypermultiplet can be put in one-to-one 
correspondence~\senstuff\ with the 243 complex deformations of ${\cal M}_1$.  
If we think of F-theory on ${\cal M}_1$ as a limit of IIA as in~\vafamori,
then the second half of every hypermultiplet comes from RR-scalars.
The origin of these scalars is less clear in F-theory.
Our expectation is that they should all
originate from Wilson lines in those portions of moduli space where all 
7--branes lie on non-degenerate surfaces.  Unfortunately, when
matching with the $T^4/\ZZ_2$ orientifolds, we encounter just such
degenerating surfaces.   $O7_2$ planes crossing at right angles can be
thought of as a degenerate complex hyperboloid; 
the two-form fluxes at their intersection are most likely 
associated with a Wilson line around this collapsed one--cycle of the 
hyperboloid~\foot{We thank C. Vafa for pointing this out to us.}.

	Now that we understand the closed string sectors of the 5--9
and 7--7' picture for the $\ZZ_2$ orientifolds, let us take a closer look at
the open string sector.  In the 5--9 picture when two
5--branes~\foot{By 5-branes we mean a pair of D5-branes glued together
by the orientifold group action.  In general, we use the notation
n-brane for an irreducible collection of Dn-branes, each of which is
labeled by exactly one Chan-Paton factor.}
coincide, the gauge group is enhanced from $USp(2)\times USp(2)$ to
$USp(4)$~\foot{Our conventions for $Sp$ groups are as follows.  We will denote
the rank {\bf n} gauge algebra $Sp(n)$, but refer to the gauge group as
$USp(2n)$ since we realize it in terms of $2n \times 2n$ unitary matrices.}.  
How does this work for 7--branes?  The analog of placing 
5--branes together is again to have 7--branes coincide, but also to match their
Wilson lines along the 67 directions. When the two $USp(2)$ Wilson
lines are equal,   
the overall Wilson line is proportional to the invariant anti-symmetric 
matrix of $USp(4)$ and so does not break down the group.  Thus as expected, the
combination of equal 89 coordinates and 67 Wilson lines for 7--branes gives the
same gauge group as for overlapping 5--branes in the 5-9 picture. This
can be generalized for $USp(2)^n$ getting enhanced to $USp(2n)$ with
$n$ overlapping 7--branes. 

	 Let us use this analysis for a further understanding of the 7--brane
gauge enhancement patterns.  When $n$ 5--branes coincide with a
$\ZZ_2$ orientifold point whose three metric blow-up modes are set to
zero, there is a gauge group enhancement to $SU(2n)$.  In the 7--7'
picture, this corresponds to placing 7--branes on an $O7_2$ plane.
Note that each of these ``planes''  has 4 fixed points on its
world--volume.  For specific values 
of the Wilson lines, we can tune the metric blow-up mode doublets, along 
with one combination of the two-form ``theta angles'', at each of the fixed
points to get $SU(2n)$.  There will be four values of the Wilson lines were
we can do this, each of which needs a different linear combination of the
$SU(2)_R$ triplet of twisted sector fields at the four fixed points to be set 
to zero.  For a comparison of this orientifold with F-theory, we would
like to set the Wilson lines and two-form fluxes to zero, since they
are not described directly as complex deformations in F-theory.  This
is not entirely possible, however, as the orientifold typically has
non-zero two--form fluxes.  

\lref\sagtensor{A. Sagnotti, {\it A Note on the Green--Schwarz Mechanism
in Open--String Theories}, hep-th/9210127, \plb{294} (1992) 
196-203.}

\lref\bpsag{M. Bianchi, G. Pradisi, A. Sagnotti, {\it Toroidal 
Compactification and Symmetry Breaking in Open String Theories}, \npb{376}
(1992) 365-386}

\lref\sensethi{A. Sen, S. Sethi, {\it The Mirror Transform of Type~I Vacua in
Six Dimensions}, hep-th/9703157, \npb{499} (1997) 45-54.}
	
\lref\edvect{E. Witten, {\it Toroidal Compactification Without Vector
Structure}, hep-th/9712028.}

\lref\aspinwall{P. S. Aspinwall, {\it Enhanced Gauge
Symmetries and K3 Surfaces}, hep-th/9507012, \plb{357} (1995) 329.}

In the 5-9 picture, the GP $\ZZ_2$ 
models are Type I compactifications with no vector
structure~\refs{\berkooz,\edvect}.   As such they have discrete
NS--NS two--form  
``theta angles''~\refs{\bpsag,\sensethi} 
which come from the fact that the NS--NS two--form takes half-integer 
values in $H_2(K3,\IR)$.  In fact, it is natural to expect these 
``theta angles'' to have non-zero value.  We
can think of the GP $\ZZ_2$ orientifold in terms of an $\Omega$ projection on
the $T^4/\ZZ_2$ orbifold compactification of IIB string theory.  But we 
know what the IIB orbifold
``theta angles'' are~\aspinwall.  They are precisely the values needed
to obstruct  
``vector structure'' in the Type~I compactification. 
In the 7-7' picture, these ``theta angles'' are no longer constrained to
take discrete values, yet their background value still starts out non-zero.
This makes it possible to find special gauge group enhancements in
the moduli space of ${\cal M}_1$ which are not immediately obvious in the
orientifold picture~\senstuff.  Further details of the
duality of  
GP $\ZZ_2$ models with F-theory on the Calabi--Yau 3--fold ${\cal M}_1$
will be left to sections 3 and 5.  We will now discuss the $\ZZ_4$ 
orientifold as it is the focus of our exploration of the 
relationship between F-theory and orientifolds.

\subsec{$\ZZ_4$ Basics}

	In the previous section we used the $\ZZ_2$ orientifold to
develop the tools 
necessary for understanding IIB orientifolds with 7-7' branes.  In addition,
we understand how data about their gauge enhancement patterns can be 
extracted from previously known result for their 5-9 duals. d.  For
the $\ZZ_4$  orientifold, the
orientifold group, after T--duality, is the product:
\eqn\omegazfour{
\{1,\Omega (-1)^{F_L} R_{67}, \Omega (-1)^{F_L} R_{89}, R_{6789}\} 
\times \{1,\alpha_4\}.
}
Here $\alpha_4$ has the 
following action:
\eqn\zfour{
\alpha_4: 
\cases{z_1 & = $X^6+iX^7 \rightarrow e^{\pi i \over 2}z_1$, \cr
z_2 & = $X^8+iX^9 \rightarrow e^{-{\pi i\over 2}}z_2$,
}
}
This form illustrates how this orientifold is related to the 
$\ZZ_2$ orientifold.  We take the $\ZZ_2$ orientifold, 
fix $\tau=i$ 
for the 67 and 89 tori, and then gauge a further $\ZZ_2$
\foot{Here, $\alpha_4$ can be thought of as a $\ZZ_2$ action on this 
orientifold
with a $\ZZ_2 \times \ZZ_2$ structure.  Of course, $\alpha_4$ has a $\ZZ_4$
action on the 
covering space. That is why we end up with an orientifold with a $\ZZ_2 \times 
\ZZ_4$ structure, where the $\ZZ_2$ comes from $\Omega R_{67}$.}.

If we look at one torus, say along the 67 directions, in the $\ZZ_2$ 
orientifold
it will be wrapped by the 89 $O7_2$ planes and 7--branes, but will also 
have localized on it four $O7_2$ planes and eight 7'--branes.  The $z_1$
coordinates of the $O7_2$ planes are:
\eqn\otwoplanes{
z_1 = {1 \over 2},\quad {i \over 2},\quad 0,\quad {(i+1) \over 2}
}
Under the action of $\alpha_4$ the first two $O7_2$ planes are
identified and the last two become $O7_4$ planes.  The eight 7'--branes
pair up to become four 7'--branes. Similarly, $\a_4$ leaves the other
torus with one $O7_2$ plane, two $O7_4$ planes and four 7--branes. 

	Where an $O7_2$ plane intersects an $O7_2$ plane or an $O7_4$
plane we have an $A_1$ singularity.  Furthermore, we also get an
$A_3$ singularity where an $O7_4$ plane intersects an $O7_4$ plane.
Note that because $\alpha_4$ acts on both $z_1$ and $z_2$  
at the same time, an $O7_2$ plane will intersect another $O7_2$ plane at two
distinct points.  Thus we have six $A_1$ and four $A_3$ singularities
as expected for the   
$T^4/\ZZ_4$ orbifold limit of $K3$.

	From Gauss' Law, it is clear that the $O7_4$ planes will have half 
the charge of the $O7_2$ planes. Thus in each torus we can cancel all the 
charge locally by placing one 7--brane (for each torus) at each of the 
$O7_4$ planes, and the  remaining two at the $O7_2$ plane.  In this situation 
there will be no dilaton gradient. Therefore we expect that the orientifold 
picture will be exact and any corrections that we get from an F-theory 
analysis should be trivial at this point in parameter space.  
Perturbations of the physics near the $O7_2$ plane has already been
carried out in the 
context of the $\ZZ_2$ orientifold~\senstuff. The key, 
then, is to understand what happens near an $O7_4$ plane.  Before we start  
analyzing the situation using F-theory, we will first go over the spectra 
predicted from the orientifold analysis~\clifferic.

	In the bulk a 7--brane carries on it an $SU(2)$ gauge group.  It 
intersects with 
each of the four perpendicular 7'--branes in two distinct points where 7-7' 
strings in a $({\bf 2,2})$ representation live.  Thus, as expected,
the low energy  
excitations on the brane correspond to an $SU(2)$ gauge group with 
16 fields in the fundamental representation.  When $n$
7--branes coalesce, we get an $USp(2n)$ gauge group with 16 fundamentals and
one antisymmetric.
At an $O7_2$ plane, this gauge group will get enhanced to $SU(2n)$ with 16
fundamental and two antisymmetrics.  Finally, at an $O7_4$ plane,
$n$ 7--branes will give an $SU(2n) \times SU(2n)$ gauge group with eight 
fundamentals, one antisymmetric in each $SU(2n)$ as well
as one $({\bf 2n,2n})$ representation.  To get a better handle on how 
this last pattern of
gauge enhancement is affected by twisted closed string sectors, let us first 
look at the  dual 5-9 picture.

\lref\joetensors{J. Polchinski, {\it Tensors from K3 Orientifolds},
hep-th/9606165, \prd{55} (1997) 6423 .}

\lref\gibhawk{G. W. Gibbons and S. W. Hawking, {\sl `Gravitational
Multi--Instantons'}, Phys. Lett. {\bf B78} (1978) 430.}
\lref\eghan{T. Eguchi and A. J. Hanson, {\sl`Asymptotically Flat 
Self--Dual Solutions to Euclidean Gravity'}, Phys. Lett. {\bf B74} (1978) 249.}

\def\y{{\bf y}}
\def\A{{\bf A}}
\topinsert{
\vskip0.5cm
\centerline{\epsfxsize=4.0in
\epsfbox{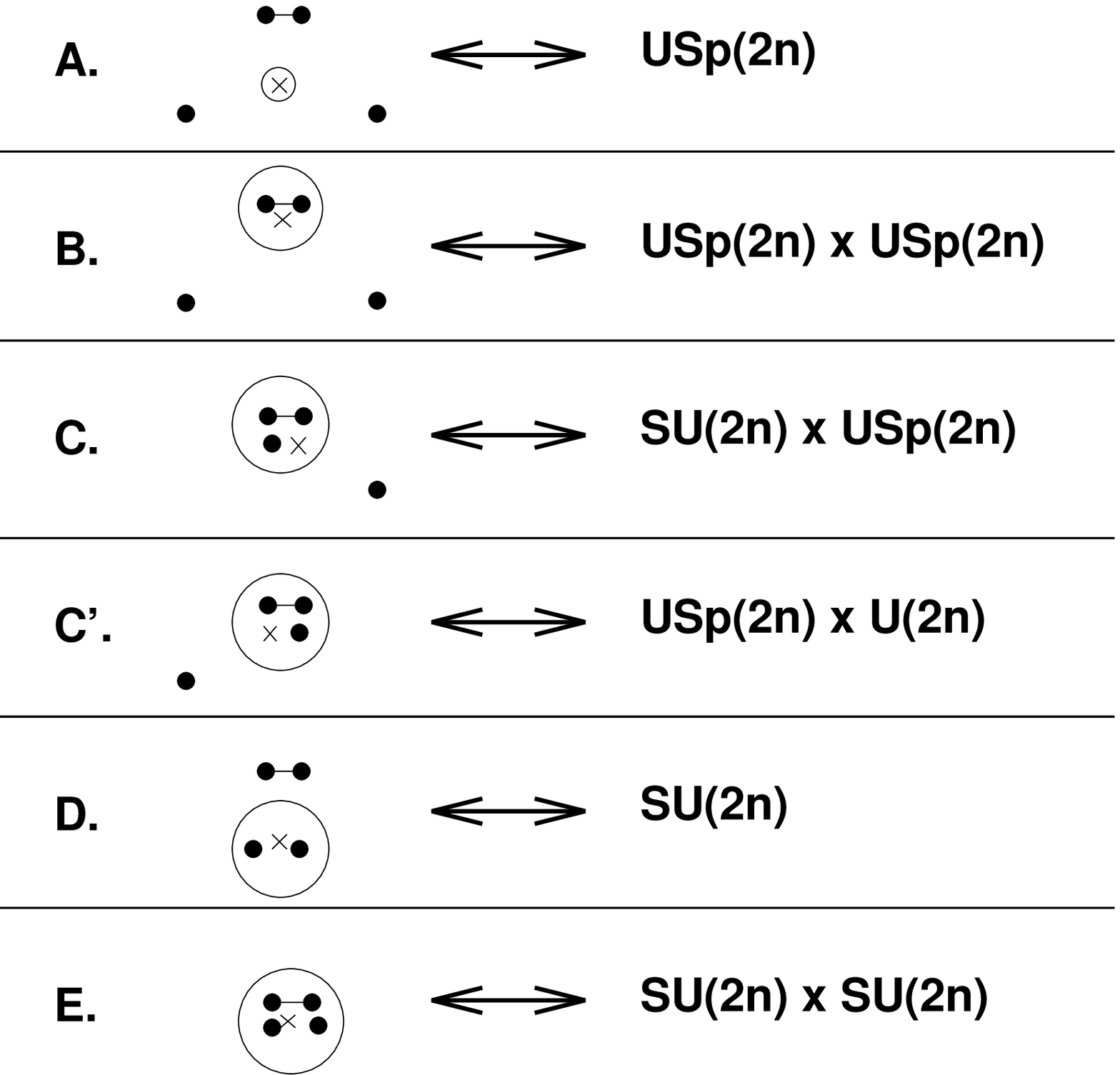}}
\vskip0.5cm}
\noindent{{\bf Figure  1.} \sl The gauge enhancement patterns for the
$\A_3$ orientifold point.  The $\y_i$ are represented by dots, and the 
"frozen" two--cycle $|\y_2 - \y_1|$ is represented by a line.  The $\times$
represents the position of a D5--brane, the circles are drawn to indicate 
overlapping objects.}\endinsert

	If we take a 5--brane $USp(2n)$ unit and place it at a $\ZZ_4$ 
orientifold point, we will get an $SU(2n) \times SU(2n)$ gauge 
group~\clifferic.  By blowing up the orientifold point and solving the
D-flatness conditions, we can higgs to a variety of  gauge
groups~\ref\gimon{E. G. Gimon, work in progress.}.  The $\ZZ_4$ orientifold 
point has two blow-up modes.  It consist of an $A_3$ singularity with one 
"frozen" cycle~\joetensors.  If we take the standard 
form~\refs{\gibhawk,\eghan} for the $A_3$ metric we have:
\eqn\Ametric{\eqalign{ds^2&=V^{-1}(dt-\A\cdot d\y)^2+V d\y\cdot d\y,\cr
{\rm where}\quad V&=\sum_{i=0}^{3}{1\over |\y-\y_i|}\quad
{\rm and}\quad \nabla V=\nabla\times\A.}}
We now set $\y_1 = \y_2$, leaving us with a pair of two--cycles
whose sizes are determined by $|\y_1 - \y_0|$ and $|\y_3 - \y_2|$.
Scenarios for the various intermediate 
gauge groups are shown in fig.~1.

  	To summarize, when $\y_0$ and $\y_3$ have generic values (case A) we 
are left with a special $\ZZ_2'$ orientifold 
point~\refs{\clifferic,\joetensors}.  When the $USp(2n)$ unit is placed on this
$\ZZ_2'$ point (case B) we get a $USp(2n) \times USp(2n)$ 
gauge group.  By setting one or the other of the blow-up modes to zero 
(cases C and C'), we can get $SU(2n) \times USp(2n)$ and 
$USp(2n) \times SU(2n)$ respectively.  We can also tune the two blow-up modes 
($\y_3 = \y_0$) so as 
to create a regular GP $\ZZ_2$  
orientifold point separate from the $\ZZ_2'$ point.  Placing 5--branes
there will give us an $SU(2n)$ gauge group (case D).  In the most
singular limit, the original orientifold point, we recover the full
$SU(2n) \times SU(2n)$ gauge group.

The twisted sectors of
a $\ZZ_4$ orientifold point also include an ${\cal N}=1$ tensor
multiplet~\clifferic. This type of multiplet contains a single scalar
whose vev, we expect, will control the relative couplings of any
product gauge group associated with placing 5-branes on the
singularity. Combined with the tensor scalars from the other $\ZZ_4$
orientifold points, it will  also control the relative couplings for
the 9-brane gauge group.

	Using the tools from section~2.3 we can now switch to the
7-7' picture.   
The 5-9 analysis above implies that a cluster  of $n$ 7--branes will enhance
its world-volume gauge group from $USp(2n)$ to
$SU(2n) \times SU(2n)$ when it is placed on an $O7_4$ plane, provided
the correct linear 
combination of blow-up modes has been set to zero.  Turning on the appropriate
modes will then yield the subgroups listed above.

In the 7-7' picture, each ${\cal N}=1$ tensor multiplet gets its
scalar component from metric deformations. To be more precise,
this scalar component  will control the area of a
two-cycle. Naively, shrinking this two-cycle should lead
to the appearance of a light non-critical ${\cal N}=1$
string. However, this can not be the case as the tensor scalar vevs in the
orientifold  are naturally zero.  The conformal field theory
describing the orientifold does not exhibit any of the singular
behaviour associated with non-critical strings. In fact, we can only
get this type of  behaviour if we tune ``theta angles'' to zero. This
implies that the ${\cal N}=1$ tensor multiplet, along with one of the
hypermultiplets which modifies the ``theta angles'', couples to an
${\cal N}=2$ non-critical string. Because this happens for any value
of the string coupling, we expect that in the F-theory description of
the $\ZZ_4$ orientifold, the relevant two-cycles will have
self-intersection number -2 (see ref.~\wittenphase).

\newsec{F-theory Interpretation of the $T^4/\ZZ_2$ Orientifold}

Let us start by briefly reviewing  the F-theory realization of the GP $\ZZ_2$
orientifold. (For more details, we refer
to~\senstuff.) Rather than
starting from the orientifold and trying to derive the relevant
F-theory compactification we will simply give the Calabi--Yau manifold,
${\cal M}_1$,
on which F-theory is compactified and point out the
correspondence with the $\ZZ_2$ model. 
This review will be useful as we go to the F-theory vacuum
relevant for the $\ZZ_4$  orientifold as it is realized in
terms of a $\ZZ_2$ orbifold of F-theory on ${\cal M}_1$.

Consider the following elliptically fibered Calabi--Yau hypersurface, ${\cal
M}_1$, with base $\IP^1\times \IP^1$, 
\eqn\weierstrass{
 Y^2 = X^3 + X f(z,w) + g(z,w)
}
where $(z,w)$ are the coordinates on the base, and $f(z,w)$ and $g(z,w)$
are of bi-degree 8 and 12 respectively~\vafamori; the above equation defines
a torus for fixed $(z,w)$. The
fiber has singularities where the discriminant, $\Delta = 4 f^3 + 27 g^2$,
vanishes. 
Since we want to compare this F-theory
vacuum with the $\ZZ_2$ orientifold, which is a type IIB
compactification, we are interested in finding configurations for which
the string coupling is constant. In
F-theory this coupling, complexified as
$a+ie^{-\phi}$, is identified with the $\t$ parameter of the
torus in eq.~\weierstrass.
By studying the modular invariant $j(\t)$-function we can deduce
properties of $\t$ and hence of the coupling constant.
In terms of $f$ and $g$ we have that $j(\t)$ can be expressed as~\senstuff,
\eqn\jfunc{
j(\t) = { 4\cdot (24 f)^3\over 4 f^3 + 27 g^2}\, .
}
Thus, constant $j(\t)$ implies that $f^3/g^2={\rm const}$. We can
satisfy this condition by choosing~\seneightd,
\eqn\fgconst{
\eqalign{f & = (\prod_{i=1}^4 (z-z_i)(w-w_i))^2 \cr
g & = (\prod_{i=1}^4 (z-z_i)(w-w_i))^3}
}
Let us compare this with the orientifold picture. 
This choice of $f$ and $g$ corresponds
to a configuration clustering  the 24 D7--branes in groups of six
around the four points $z_i$ (and similarly for the D7--branes in the
$w$-plane).  If we count two D7--branes in each of the clusters as making
up one orientifold plane~\seneightd, this leaves 16 D7--branes on each
$\IP^1$,  
just what we expect from the $\ZZ_2$ model.  Unfortunately, we do not have an
exact correspondence.  
In F-theory, the configuration above describes colliding $D_4$
singularities, a situation which heralds the presence of ``tensionless
strings''~\ref\collide{M. Bershadsky, A. Johansen, {\it Colliding 
Singularities in F-theory and Phase Transitions}, hep-th/9610111, \npb{489}
(1997) 122-138}.  On the orientifold side, the $\ZZ_2$ model has non-zero 
``theta-angles'' which drive the theory away from this critical point.  

	Let us consider, therefore, a more generic situation from the 
orientifold point of view.  For this we move sets of 7--branes off the
orientifold planes.  Theoretically, we could move off sixteen independent 
D7--branes in each $\IP^1$~\foot{We can think of $T^2/\ZZ_2$ as $\IP^1$, the
16 D7--branes come from 32 D7--branes on the covering space}.
The orientifold group action, however, pairs these up into eight
groups of two,  
each contributing an $SU(2)$ gauge group (see the discussion in
section~2.2).  We therefore would like  
to find a choice of $f,g$ corresponding to this $(SU(2)\times SU(2)')^8$ and
containing objects which become orientifold planes at weak coupling. 

\lref\bershad{M. Bershadsky et al, {\it
Geometric Singularities and Enhanced Gauge Symmetries}, hep-th/9605200,
\npb{481} (1996) 215.}
The F-theory gauge groups are given in terms of the singularities
of the elliptic fiber.  
This singularity structure is encoded in the behavior of
$\Delta, f,g$~\bershad. 
In order to obtain the $(SU(2)\times SU(2)')^8$, we need eight
$A_1$ singularities in each of the $z,w$ planes. This implies 
\eqn\deltasutwoa{
\Delta \sim \prod_{i=1}^8 (z- z_i)^2 (w- w_i)^2,
}
with $f,g$ non-vanishing as $z\to z_i,\, w\to w_i$.
This is obtained by making the following choices~\senstuff,
\eqn\fgsutwo{
\eqalign{f &= \eta - 3 h^2 \cr
g &= h(\eta - 2 h^2) \cr
\eta &= C \prod_{i=1}^8 (z-z_i)(w-w_i) \cr
h &= \prod_{i=1}^4 (z - \tilde z_i)(w - \tilde w_i)}
}
for which 
\eqn\deltasutwob{
\Delta = C^2 \prod_{i=1}^8 (z- z_i)^2 (w- w_i)^2 (4\eta - 9 h^2)\, .
}
This choice of $h$ and $\eta$ is motivated by the need to recover
an orientifold at weak coupling.  Taking $C\to 0$ sends $j(\l)\to \infty$
almost everywhere, see eq.~\jfunc.   This implies that up to an
$SL(2,\ZZ)$ transformation 
we have weak coupling almost everywhere.  Note that the last factor in
eq.~\deltasutwob\ yields pairs of 7--branes  centered about
the zeros of $h$ and separated by a distance
of order $C$. These are the  orientifold planes~\senstuff.
Deformations  
of $h$ are mapped to the blow-up modes of the type IIB orientifold. 
By moving the $z_i,w_i$ around (\ie\ the locations of the 7--branes in
the orientifold picture), we can enhance the symmetry further.  
In particular, if $n$ of the $z_i$ coincide one gets an 
$Sp(n)$ algebra, and if it happens at the orientifold plane there is the
possibility of further enhancement~\senstuff.

\lref\gopakumar{R. Gopakumar, S. Mukhi, 
{\it Orbifold and Orientifold Compactifications 
of F-theory and M-theory to Six and Four Dimensions}, hep-th/9607057,
Nucl. Phys B479 (1996) 260-284.}

\lref\blum{
J. Blum, A. Zaffaroni, {\it An Orientifold From F Theory}, hep-th/9607019,
Phys.Lett. B387 (1996) 71
\semi J. Blum, {\it F-Theory Orientifolds, M-theory Orientifolds, and
Twisted Strings}, hep-th/9608053, Nucl. Phys. B486 (1997) 34-48.}

\newsec{F-theory Interpretation of the  $T^4/\ZZ_4$ Orientifold}
Let us next turn to
the type IIB orientifold $T^4/\ZZ_4$ and in particular the realization
(and extension) of the orientifold in terms of F-theory. As above, we will
mainly discuss the F-theory compactification and where appropriate,
compare with the $\ZZ_4$ orientifold. In section~2 we showed that one
can construct a T-dual version of the original Gimon-Johnson 
$\ZZ_4$ orientifold as a $\ZZ_2$ orbifold of the GP-model. Thus we
are  naturally lead to build the corresponding F-theory vacuum as a
$\ZZ_2$ orbifold of ${\cal M}_1$~\foot{A general treatment of
F-theory on orbifolds has yet to be done. 
See~\refs{\vafamorii,\blum,\gopakumar} for some more examples of such 
constructions.}.

\subsec{Construction of the F-theory orbifold}
Starting with our elliptically fibered Calabi--Yau with base $\IP^1\times
\IP^1$ we construct an orbifold, $\overline{\cal M}_2={\cal
M}_1/\ZZ_2$, using the $\ZZ_2$ quotient $(z,w) \to (-z,-w)$.  
There are four $\ZZ_2$ fixed points in the base, or $A_1$
singularities, and hence four fixed tori in
the Calabi--Yau manifold. Each fixed torus will contribute one K\"ahler
deformation, from a 2-cycle
living on a $\IP^1$ of the blown-up torus, and one complex structure
deformation, from a 3-cycle built out of a family of $\IP^1$s over a 
1-cycle of the torus. As we will show in section~4.3, there are 123 complex
deformations invariant under the $\ZZ_2$ quotient. 
If we extend our notion $\overline{\cal M}_2$, beyond its strict definition as 
an orbifold with unresolved fixed tori, to the surface were the singularities
are slightly resolved, then the total number of complex structure deformations
is $h_{2,1}(\overline{\cal M}_2)=123+4=127$. 
Similarly, the four new K\"ahler deformations join the three inherited
from ${\cal M}_1$ to give $h_{1,1}(\overline{\cal M}_2)=3+4=7$.

\lref\louis{J. Louis, J. Sonnenschein, S. Theisen, S. Yankielowicz,
{\it Non-Perturbative Properties of Heterotic String Vacua Compactified on
     ${K3\times T^2}$}, hep-th/9606049, \npb{480} (1996) 185.}

\lref\candelas{P. Candelas, E. Perevalov, G. Rajesh, {\it F-Theory
Duals of Nonperturbative Heterotic E8xE8 Vacua in Six Dimensions},
hep-th/9606133, \npb{502} (1997) 613.}

To understand where the fixed tori come from we study the definition
of the manifold as a hypersurface in a toric 
variety. (For a  more detailed discussion of toric geometry in
relation to F-theory, see for
example~\refs{\vafamori,\vafamorii,\louis,\candelas.})  
A hypersurface in
(weighted) projective space is defined using a scaling
relation (also known as a $\IC^*$ action),
\eqn\cstar{
x_i \to \l^{k_i}\, ,\quad p(x_i)\to \l^d p(x_i),
}
on a defining polynomial, $p(x_i)=0$, of degree $d=\sum_i k_i$. In a
toric variety, there are more coordinates and hence a 
larger number of scaling relations. In particular, the elliptic fibration over
a base $\IP^1\times \IP^1$, ${\cal M}_1$, can be described as a  
hypersurface in a toric variety  with seven homogeneous coordinates
and three $\IC^*$ actions~\vafamori;
\eqn\cstarpxp{
(s,t,u,v,X,Y,Z)\to 
(\l_1 s, \l_1 t, \l_2 u, \l_2 v,(\l_1\l_2)^4\l_3^2 X,(\l_1\l_2)^6
\l_3^3 Y,\l_3 Z).
}
By a rescaling of $\l_3$ we set $Z=1$. 
We define our  inhomogeneous coordinates as $z=s/t$, $w=u/v$.
By setting $|\l_i|=1$, this implies in particular 
the existence of three discrete identifications, each one associated with the
$\IC^*$ actions above. This can be written in a
more compact notation as
\eqn\giaction{
g_1:\,  (\ZZ_{12}:0,0,1,1,4,6,0)\, ,\quad 
g_2:\,  (\ZZ_{12}:1,1,0,0,4,6,0)\, ,\quad 
g_3:\,  (\ZZ_6: 0,0,0,0,2,3,1)\, ,}
where $(\ZZ_d: a_1,...,a_7)$ implies 
\eqn\giexpl{
(s,...,z)\to (\a^{a_1}s,...,\a^{a_7}z), \quad \sum_i a_i = 0\, ({\rm mod}\, d),
\quad \a^d=1.
}
In this notation we can express our $\ZZ_2$ action as follows,
$\tilde g: (\ZZ_2:1,0,1,0,0,0,0)$. Combining the various actions we then find
the following fixed points 
\eqn\gifixed{
\twoeqsalign{\tilde g:\,  &    (\ZZ_2:1,0,1,0,0,0,0)\, ,\quad s=u=0, \quad
&g_1^6 \tilde g:\,  (\ZZ_2:0,1,1,0,0,0,0)\, ,\quad &t=u=0, \cr
g_2^6 \tilde g:\,  & (\ZZ_2:1,0,0,1,0,0,0)\, ,\quad s=v=0, \quad
&(g_1 g_2)^6 \tilde g:\,  (\ZZ_2:0,1,0,1,0,0,0)\, , \quad & t=v=0. \cr}
}
{}From the definition of the inhomogeneous coordinates, $(z,w)$ we see
that the four $\ZZ_2$ fixed points in the base are given by 
$(z,w)=\{(0,0),(0,\infty),(\infty,0),(\infty,\infty)\}$.

\subsec{Comparing F-theory with the type IIB orientifold}

How does the spectrum of this F-theory vacuum, $\overline{\cal M}_2$,
compare with the type IIB $T^4/\ZZ_4$ orientifold? As shown
in~\clifferictwo, it is possible to Higgs the gauge
symmetry completely in the orientifold theory. This gives us a theory with 
$n_H=128$ hypermultiplets, $n_T=1+4=5$ tensor multiplets, and
$n_V=0$ vector multiplets. Except for the ${\cal N}=1$
tensor multiplet inherited from six-dimensional supergravity, the
tensor multiplets act like
${\cal N}=2$ multiplets formed by one ${\cal N}=1$ tensor multiplet
and one ${\cal N}=1$ 
hypermultiplet. Thus, four of the 128 hypermultiplets are on a
different footing. 

For F-theory on $\overline{\cal M}_2$, 
a generic choice of complex structure gives no gauge enhancements,
and hence $n_V=0$. Furthermore, the number of complex structure
deformations is related to the number of hypermultiplets by $n_H=h_{2,1}+1$,
where the extra contribution comes from the volume of the
Calabi--Yau manifold~\vafa. Since $h_{2,1}=127$, this agrees with the
orientifold analysis. 
The total number of tensor multiplets is given by $n_T=h_{1,1}({\rm
Base})-1$~\vafamorii.  
Since $h_{1,1}({\rm Base})=6$, we find that the spectrum of F-theory
compactified on ${\cal M}_2$ is in agreement with that of the
type IIB $T^4/\ZZ_4$ orientifold. 

   This agreement becomes even more natural, if we study the correspondence 
between the tensor multiplets and the fixed points in the base of 
$\overline{\cal M}_2$ in more detail.
In resolving an $A_1$ singularity, we replace the $\ZZ_2$ fixed
point by a two-cycle, whose self-intersection number is -2. This is
a different phenomenon than that of blowing up a regular
point in the base, in which we obtain an exceptional  divisor with
self-intersection 
number -1. In particular, there 
is a non-toric complex structure deformation associated to the $A_1$
singularity, by which the singularity can be 
deformed. Thus, we have an effective
${\cal N}=2$ tensor multiplet containing an ${\cal N}=1$ hypermultiplet in
addition to the usual ${\cal N}=1$ tensor. In all, resolving the four $A_1$s
give us four ${\cal N}=2$ tensor multiplets just as in the type IIB
$T^4/\ZZ_4$ orientifold. (For a similar
discussion of $A_1$ singularities in F-theory,
see~\refs{\vafamori,\wittenphase}.)

\subsec{Blow-ups, deformations and gauge enhancement}

Having shown that the spectra agree, we now turn to a more detailed
comparison between the models. In particular, we want to study how the
fixed point deformations and the gauge enhancement in the type IIB 
orientifold arise on
the F-theory side.  In section 2.3, we described the generic configuration
of the $T^4/\ZZ_4$ orientifold.  It had four 7--branes and four
7'--branes, each with an $SU(2)$ gauge group on its world--volume.  In this
section we show how the complex structure deformations of $\overline{\cal M}_2$
can be tuned to get appropriate $O7$ planes~\foot{We will use $O7$ plane to
refer to both the $O7_2$ and $O7_4$ orientifold planes.} along with the branes
carrying the $SU(2)^4\times SU(2)'^4$
gauge group.  We will then sketch how further tuning can place these branes on
$O7_4$ planes and how the gauge enhancement patterns in fig. 1 can take place.  
We leave most of the technical details to Appendix A.

To tune the complex structure of $\overline{\cal M}_2$, we
first need to determine how it descends from that of ${\cal M}_1$,
We keep only deformations left invariant by the $\ZZ_2$ orbifold action. In 
terms of the defining equation~\weierstrass, we restrict $f,g$ to terms which 
are $\ZZ_2$ invariant. This reduces the number of binomials from 81 and 169 to
41 and 85 for $f(z,w)$ and $g(z,w)$, respectively. As before, we can 
rescale the defining
equation by an overall factor, which removes one degree of
freedom. Although the $SL(2,\IC)$ reparameterization of each of the
$\IP^1$ has been broken by the $\ZZ_2$ action there is one ``rescaling''
that can be done.  A one-parameter subgroup of the original $SL(2,\IC)$
leaves $\t =i$ invariant.  We are
thus left with 123 parameters, were we have momentarily neglected 
the additional contribution to the number of complex structure
deformations from the four $\ZZ_2$ fixed points in the base.

In order to have an $(SU(2)\times SU(2)')^4$ and respect the quotient 
symmetry, we take
\eqn\hetazfour{
\eqalign{\eta &= C \prod_{i=1}^4 (z^2 - z_i^2)(w^2 - w_i^2)\cr
h & = Q(z,w) (z^2 - \tilde z^2)(w^2 - \tilde w^2)}
}
where $Q(z,w)$ is of bi-degree two, and invariant under the $\ZZ_2$ action. 
This form for $\eta$ gives, after identification, the four 7--branes and
7'--branes we require.  As was the case in section 3, $h$ controls the $O7$ 
planes.  There are branes making up $O7_2$ planes about $z = \tilde z$ and
$w = \tilde w$.  $Q(z,w)$ is related to the $O7_4$ planes and their 
intersections.  Note that we have chosen the most generic form for $h$ 
consistent with the $\ZZ_2$ quotient.

In~\clifferic, and as discussed in section~2,
it was shown that out of the 128 hypermultiplets sixteen are associated to
closed string sector of the $T^4/\ZZ_4$. Two of them come from the
untwisted sector and the other fourteen from blow-up modes for the fixed
points. Of these, ten come from the $\ZZ_2$ twisted sector of the $\ZZ_2$ and
$\ZZ_4$ fixed points.  An additional four come from the other twisted sector
of the $\ZZ_4$ fixed points, appearing together with the four tensor multiplets
and forming four effective ${\cal N}=2$ tensor multiplets (one tensor + five 
scalars).  

Let us account for these blow-up modes in our F-theory compactification. 
The blow-up modes for the $\ZZ_4$ orientifold are encoded in
deformations of the form of $h(z,w)$ given in~\hetazfour, in analogy
with the situation for the $\ZZ_2$ orientifold. The six blow-up modes of
the $\ZZ_2$ points come from mixing the three factors of $h$ in~\hetazfour\
consistent with the F-theory $\ZZ_2$ quotient.  $Q(z,w)$ controls the one
deformation for each of the $\ZZ_4$ orientifold points.  The last four 
deformations, the ones which pair up with the tensor multiplets, come
from the non-toric deformations. 

To realize one of these non-toric deformations
we make a change of coordinates such that the $A_1$ singularity at
$z=w=0$ is given by
\eqn\transform{
a\cdot b = c^2\,,\quad {\rm where} \quad
a = z^2,\quad
b = w^2, \quad
c = zw~.
}
Then we can deform the $\ZZ_2$ quotient singularity at $a=b=c=0$
in the base of the 
elliptic Calabi--Yau,
\eqn\quotsingdef{
a\cdot b = c^2 - \l_{11}^2~.
}
The expression for the deformations of the other $A_1$ singularities
can be found in appendix~A.

Now that we understand how the various $O7$ planes and fixed point 
deformations of the $T^4/\ZZ_4$ orientifold appear in F-theory, we examine 
enhancement patterns for the 7--branes (a similar analysis holds for the
7'--branes).  When
$n$ coinciding 7--branes are located away from an $O7$ plane, we
have an $USp(2n)$ gauge 
symmetry just as in the $\ZZ_2$ orientifold. In F-theory this is
obtained by identifying $n$ of the $z_i$'s, e.g. $z_1=...=z_n$ in the
defining equation for $\overline {\cal M}_2$~\hetazfour. The discriminant
then takes the form
\eqn\deltazfoura{
\Delta \sim (z^2-z_1^2)^{2n}~,
}
which at $z = z_1$ gives an $A_{2n-1}$ 
singular elliptic fiber.  For a generic choice of $h$, this singularity is
non-split and the gauge group has an $Sp(n)$ algebra~\bershad.  
This is the F-theory description of scenario A in the $\ZZ_4$
orientifold (see fig.~1).

The matter
content can be deduced in analogy with Sen's analysis for F-theory on
${\cal M}_1$ corresponding to the 
$\ZZ_2$ orientifold~\senstuff. We know that only $n-1$ moduli are
involved in enhancing the symmetry from $USp(2)^n$ to $USp(2n)$. They
correspond to separating the $n$ 7--branes. This is done by higgsing
$USp(2n)$ using matter transforming as ${\bf
n(2n-1)-1}$ for which $Sp(n)\to Sp(1)^n$ and we are left with $(n-1)$
${\bf 1}$.  Thus, there is one ${\bf n(2n-1)-1}$ of $Sp(n)$.

In order to study the situation of $n$ 7-branes approaching an $O7_4$ 
plane we let $z_1\to 0$.  The details of this analysis can be found in
appendix~A. We find complete correspondence between the various gauge 
enhancement patters in the $\ZZ_4$ orientifold as given in fig.~1, and
F-theory 
on $\overline{\cal M}_2$.  The crux of the matter can be understood as follows.
From eq.~\transform, one can clearly see that the divisors
corresponding to $z = constant$ can also be defined by $a = constant$, except
when $z = 0$.  For this case, the corresponding divisor can be thought of, 
using eq.~\quotsingdef, as the sum of two divisors with defining equations 
$a = 0, c = \pm \l_{11}$.  When we take $z_1\to 0$, $\Delta$ now has two
divisors with $A_{2n-1}$ singularities.  This yields the expected product gauge
groups.

\newsec{F-theory on elliptic Calabi--Yau 3-folds and the dual
heterotic theory on $K3$}

\def\[{\left[}
\def\]{\right]}

In the previous section we established the relation between the type
IIB orientifold on $T^4/\ZZ_4$ and F-theory compactified on the
orbifold $\overline {\cal M}_2={\cal M}_1/\ZZ_2$. We would now like to 
understand the dual heterotic descriptions of this model. In particular, we 
are interested in the role of the heterotic $E_8\times E_8$ theory 
compactified on $K3$ with instanton embedding $(10,10)$ and four extra 
${\cal N}=1$ tensor multiplets, the conjectured dual to the
$T^4/\ZZ_4$ orientifold 
(see ref.~\clifferictwo). 
We will demonstrate that F-theory on $\overline {\cal M}_2$ is
dual to a strong coupling limit of this heterotic $E_8\times E_8$
theory on $K3$ with instanton embedding $(10,10)$. The heterotic
theory can be described as
M-theory compactified on $K3\times (S^1/\ZZ_2)$ with four M-theory
5-branes located in pairs at two 
points in the $K3$. Surprisingly, one can reformulate this strongly coupled
theory as a new weakly coupled heterotic theory.

\subsec{The $(12,12)$ instanton embedding and its F-theory dual}

The first step in our demonstration will be a review of the relationship
between the heterotic $E_8\times E_8$ string compactified on $K3$ with
instanton embedding $(12,12)$ and its dual F-theory compactification
on ${\cal M}_1$.

\lref\hw{P. Horava, E. Witten, {\it Heterotic and Type I String
Dynamics from Eleven Dimensions}, hep-th/9510209, \npb{460} (1996)
506.}

\lref\duff{M. Duff, J. Liu, R. Minasian, {\it Eleven Dimensional
Origin of String/String Duality: A One Loop Test}, hep-th/9506126,
\npb{452} (1995) 261.}
As we later will be interested in non-perturbative effects in the heterotic
$E_8\times E_8$ theory on $K3$, let us consider the corresponding
situation in terms of M-theory on $K3\times (S^1/\ZZ_2)$~\hw, (see
fig. 2 a),b)). 
The fundamental string is represented in terms of a membrane, stretching
between the ``end-of-the-world'' 9-branes. At each of the ends, the
boundary of the membrane is a string, which carries a level one $E_8$
current algebra~\hw.  The tension
of the string is proportional to the distance between the 9-branes,
the interval $S^1/\ZZ_2$ which in M-theory units we denote by $R$;
hence as $R$  decreases we obtain the weakly coupled $E_8\times E_8$ heterotic 
string.  Let us denote this string by $het_1$. 
In addition, there exists a second heterotic string 
given by an M-theory 5-brane wrapping the $K3$~\dmw. 
We will denote that string 
by $het_2$. The two heterotic strings are
related by a duality due to the electric-magnetic duality in M-theory
between membranes and 5-branes~\duff. This duality is manifested in
six dimensions by on one hand wrapping a membrane on $S^1/\ZZ_2$ and
then reducing on $K3$, and on the other hand by wrapping a 5-brane on
$K3$ and reducing it on $S^1/\ZZ_2$. In this way one can see  that the
coupling for  
$het_1$, $\l_1$, is related to the coupling for $het_2$, $\l_2$ by 
\eqn\elmagn{
\l^2_1=(\l^2_2)^{-1} \propto R/V\,;
}
where $V$ is the volume of the $K3$, and both $R$ and $V$ are expressed in
M-theory units (for more details see ref.~\dmw). We see that in,
analogy to the weakly coupled 
$het_1$ with small $R$ relative to $V$, there is a weakly 
coupled dual heterotic string
when the volume of the $K3$ becomes small relative to $R$. When $R$
and $V$ are of comparable size, neither string description is valid and 
we turn to F-theory for a better description.

\topinsert{
\vskip1.0cm
\centerline{\epsfxsize=4.5in
\epsfbox{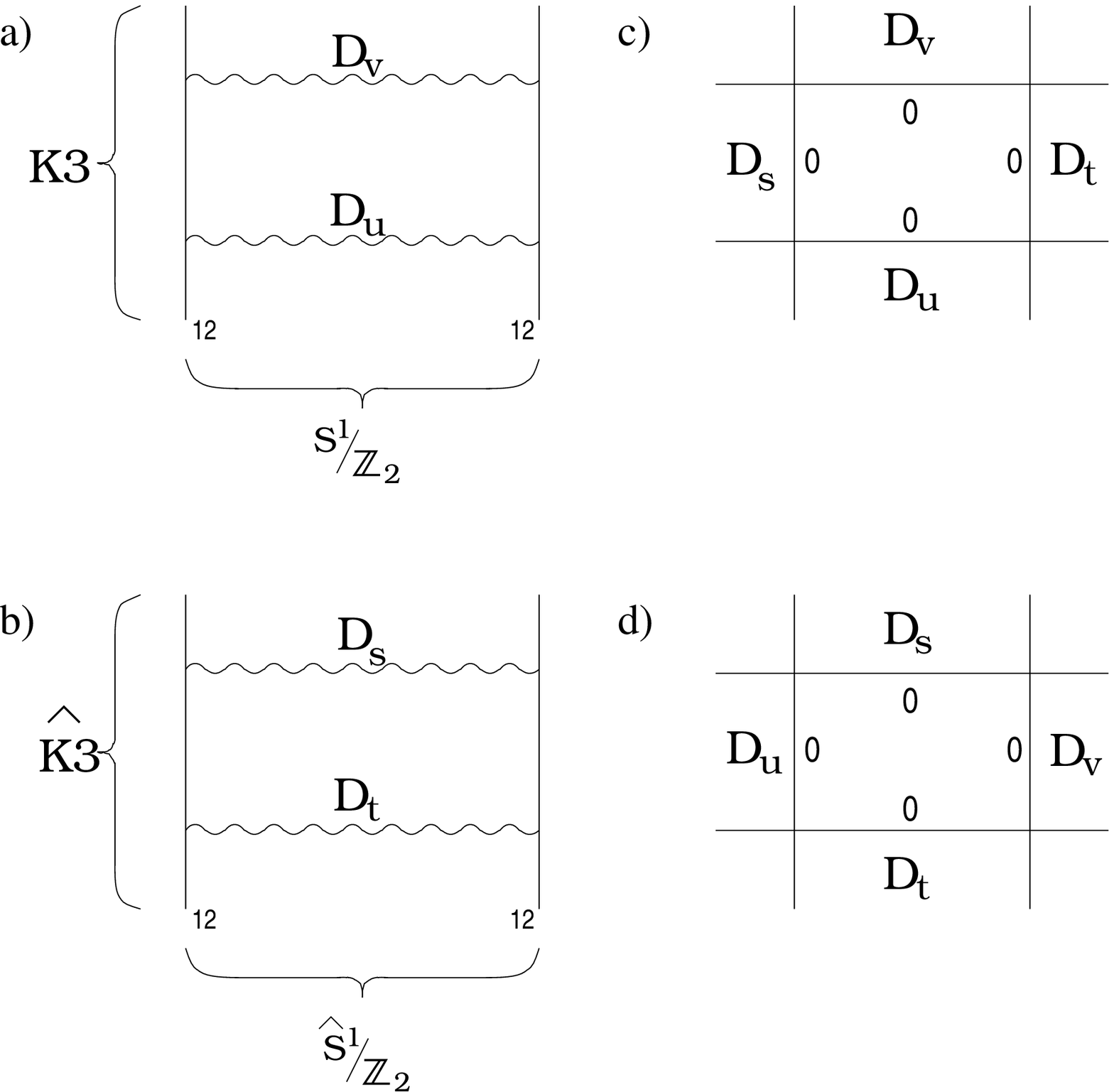}}}
{\baselineskip 10pt
\noindent{{\bf Figure  2.} \sl 

a)  Heterotic $E_8\times E_8$ string $het_1$
on $K3\times S^1/\ZZ_2$ with instanton embedding $(12,12)$.

b) Magnetic dual $E_8 \times E_8$ heterotic string $het_2$ on
dual ${\hat K3\times \hat S^1/\ZZ_2}$ background.

c) Divisors for the base $F_0 = \IP^1\times \IP^1$ of the (3,243)
Calabi--Yau dual to fig. 2a), 
where $\l_1 \propto {\rm area}(D_v)={\rm area}(D_u)$. 

d) The same base gives the F-theory dual to fig. 2b) with
 $\l_2 \propto {\rm area}(D_s)={\rm area}(D_t)$.}}\endinsert

In the dual F-theory on the elliptic ${\cal M}_1$ the picture above
is given in terms of the divisors of the base
$F_0=\IP^1\times \IP^1$ (see fig~2 b),d)). Following~\vafamori, 
we assign an area $a_d$ to the $\IP^1$ fiber of $F_0$ and
similarly an area $a_h$ to the $\IP^1$ base. In terms of the divisor classes
$\[D_s\] = \[D_t\]$ and $\[D_u\] = \[D_v\]$ ( the projective
coordinates $(s,t)$  
label one $\IP^1$ and $(u,v$) the other) these areas can be expressed as:
\eqn\areadiv{
 a_d = {\rm area}(D_s) = {\rm area} (D_t)\,,\quad
 a_h = {\rm area}(D_u) = {\rm area} (D_v)\,.
}
The area of a divisor, $D_{x_j}$,
 is computed by considering the intersection of $D_{x_j}$ with the general
K\"ahler class, $K=a_d D_v + a_h D_s$, given
that $D_{x_i} \cdot D_{x_j}=0$
unless $D_{x_i}$ and $D_{x_j}$ are neighboring divisors in which case
$D_{x_i} \cdot D_{x_j}=1$ (see fig. 2c),d)).

In order to identify this F-theory vacuum with that of the
heterotic string, we first observe that we have two types of
D-strings. They are obtained by wrapping D3-branes on elements of
either of the divisor
classes $\[D_u\]$, $\[D_s\]$. For these
D-strings, the tension is given in terms of the area of the wrapped divisor.
These D-strings are the dual heterotic strings in the six-dimensional 
heterotic theory.  This allows us to identify the six-dimensional heterotic 
coupling constant, $\l_1$, in terms of F-theory variables as~\vafamori\
\eqn\hetftheory{
\l^2_1 = a_h/a_d\,.
}
Since the overall volume of the Calabi--Yau in the context of F-theory
is a hypermultiplet, and the size of the elliptic fiber is frozen, we 
can fix the remaining ``effective'' K\"ahler parameter by choosing $a_h
a_d=1$. Thus, 
\eqn\couplingdiv{
\l_1 = a_h\,,\quad \l_2 = a_d\,.
}

We will next consider situations in which we blow up the 
base at generic points.  Much of the above analysis carries through, with the 
obvious modification of the K\"ahler class such that it now depends on the 
exceptional 
divisors from the blown-up $\IP^1$s . Also, $a_{h,d}$ will now be associated
with  new divisor classes, defined so as to contain only elements of
self-intersection number 0. There is a natural correspondence between
divisors of self-intersection number 0, or rather the 
D3-branes which wrap these divisors, and  the 
(dual) heterotic strings~\vafamori. As discussed in the introduction, the
remaining exceptional 
divisors with self-intersection number -1 and -2 correspond to ${\cal N}=1$ and
${\cal N}=2$ tensionless strings respectively, in the limit that the area of 
the given divisor goes to zero~\foot{For the case of divisors with
self-intersection number $-n$, $n > 2$, see ref.~\wittenphase}.

	Typically, some of the original divisors of $F_0$ will have modified 
self-intersection numbers after we blow up 
the base (see for example Fig. 3 c),d)).  This change is interpreted
in the dual  
heterotic strings, defined above, as follows.  If we 
consider the D-string whose coupling depends on $a_h$, it will have as a
target space an $E_8 \times E_8$ $K3$ compactification with instanton
numbers $(12+n_s,12+n_t)$ where $n_{s,t}$ are the self-intersection numbers
of the divisors $D_{s,t}$.  A similar story follows for $a_d$.  Note that for
the $(12,12)$ compactification that we have been studying this means that all
the relevant divisors have self-intersection number zero, and are thus 
appropriate for defining the areas $a_{h,d}$.

\subsec{The $(10,10)$ instanton embedding and its F-theory dual}

The next step in our demonstration will be to construct a new manifold,
${\cal M}_2$, which we will use as an F-theory compactification to produce
a dual six dimensional model for the heterotic $E_8\times E_8$ theory 
compactified on $K3$ with instanton embedding $(10,10)$. 
Using~\vafamorii, (see also \candelas) we obtain ${\cal M}_2$ by
blowing up the base  
$F_0$ of ${\cal M}_1$ at four points.  We can use ${\cal M}_2$ to represent
either of the electric-magnetic dual $(10,10)$ models, shown in fig~3a),b).
These figures represent these models in terms of M-theory on $K3\times 
S^1/\ZZ_2$, as they are inherently strongly coupled.  The crux of our 
demonstration will be to exhibit $\overline {\cal M}_2$ as a limit of 
${\cal M}_2$.
Both of these elliptic fibrations correspond to the same Calabi--Yau
three-fold 
with Hodge numbers (7,127), but the moduli space of $\overline {\cal M}_2$ is 
the subset of ${\cal M}_2$ with four unresolved $A_1$ singularities in 
base.  We 
will demonstrate how these singularities appear as we take the limit.

\topinsert{
\vskip1.0cm
\centerline{\epsfxsize=4.5in
\epsfbox{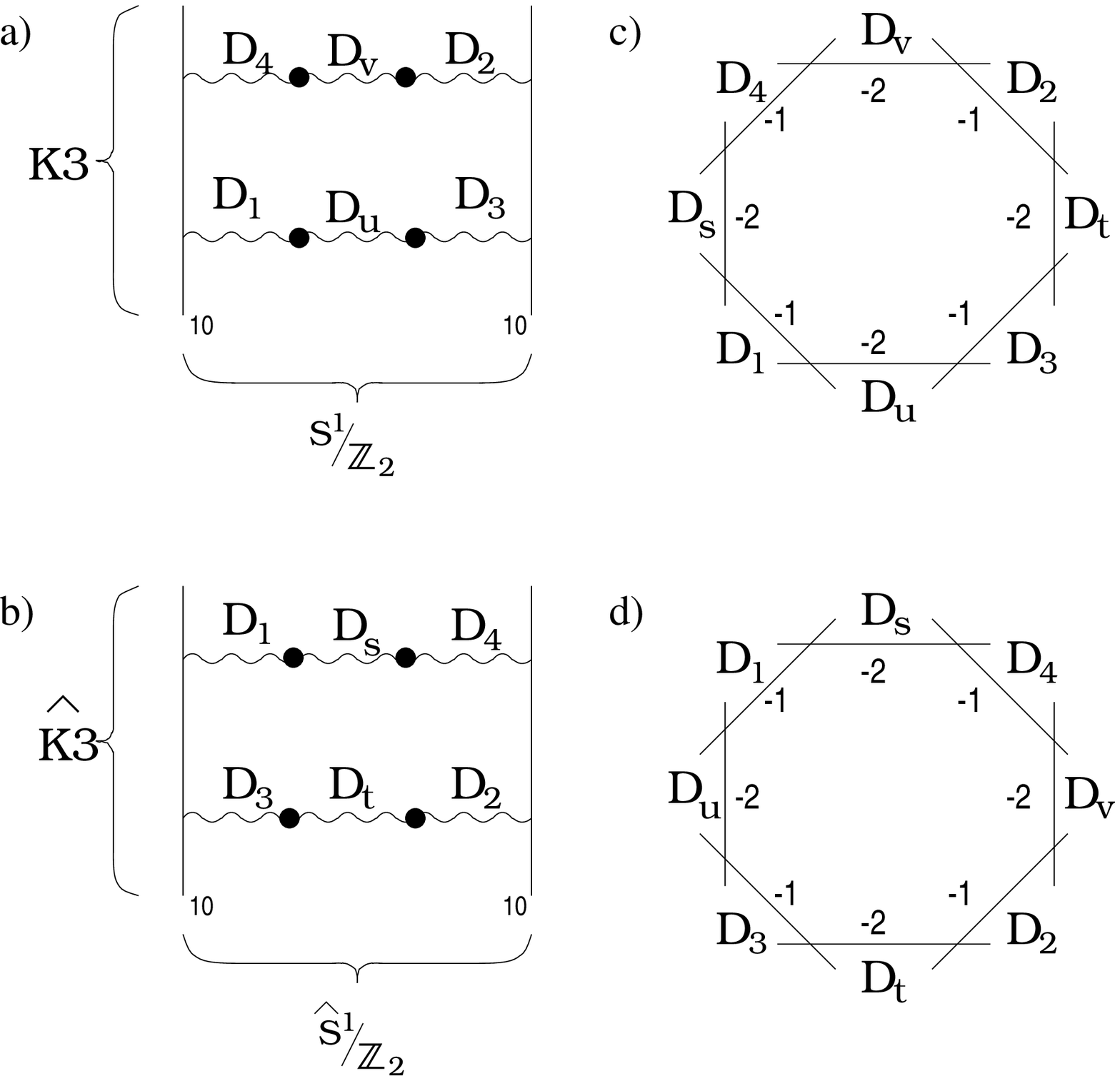}}}
{\baselineskip 10pt
\noindent{{\bf Figure  3.} \sl 

a)  Heterotic $E_8\times E_8$ string, $het_1$,
on $K3\times S^1/\ZZ_2$ w/ instanton embedding $(10,10)$.

b) Magnetic dual $E_8 \times E_8$ heterotic string, $het_2$, on
dual $\hat {K3\times S^1/\ZZ_2}$ background with $(10,10)$ embedding.

c) $F_0$ blown-up at four points, the base for ${\cal M}_2$, the
dual model for fig. 3a) where ${\l}_1~\propto~{\rm area}(D_u + D_1 + D_3)$. 

d) The same base gives the F-theory dual for fig. 3b) with
 ${\l}_2 \propto {\rm area}(D_s + D_1 + D_4)$.}}\endinsert

\lref\ruben{S. Ferrara, R. Minasian, A. Sagnotti, {\it Low-Energy
 Analysis of $M$ and $F$ Theories on Calabi--Yau Threefolds},
 hep-th/9604097, \npb{474} (1996) 323.}

\lref\sagnotti{A. Sagnotti, {\it A Note on the Green-Schwarz
 Mechanism in Open String Theories}, hep-th/9210127, \plb{294} (1992) 196.}

The heterotic $E_8\times E_8$ theory compactified  on 
$K3$ with instanton embedding $(10,10)$ can be obtained from the $(12,12)$
instanton embedding by shrinking two $E_8$ instantons on each of the
``end-of-the-world'' 9-branes.  Taking the zero-size limit of an 
$E_8$-instanton yields a new phase of the theory in which an M-theory 
5-brane is detached from the 9-brane~\refs{\ami,\sixcomments}. 
This 5-brane carries an ${\cal N}=1$ tensor multiplet whose scalar component
parameterizes the position of the 5-brane relative to the
9-brane from which it emanated.  In addition, this scalar component will
enter in the gauge kinetic
terms of the low-energy six dimensional supergravity action  in a
fashion determined by anomalies (for more details,
see refs.~\refs{\sagnotti,\ruben}).  We label the scalars corresponding
to the four 
M-theory 5-branes in the $(10,10)$ model $\phi_i,\, i= 1,\dots,4$.  Note that
these 5-branes also have coordinates in $K3$ inherited from their parent
$E_8$ small instantons.   

	{}From the generic situation for the $(10,10)$ 
model, we now want to tune parameters so as to reach a point in the
moduli space 
of this model which is connected to the $T^4/\ZZ_4$ orientifold.  One 
immediate problem arises in matching to this orientifold.  As we 
will later demonstrate, the $(10,10)$ model is only weakly coupled in the limit
where the scalars $\phi_i$ are small.  This means that we expect to have 
four light non-critical ${\cal N}=1$ strings in the theory.  We know from 
section~2, however, that the $T^4/\ZZ_4$ orientifold is in a region of moduli
space associated with four light non-critical ${\cal N}=2$ strings.  This means
that if the $(10,10)$ model is the dual of this orientifold, it must be 
strongly coupled.  In order to describe this let us turn to F-theory.

	To study the heterotic $(10,10)$ model we start with the
F-theory description of the heterotic 
$(12,12)$ model with coupling $\l_1 = a_h$, and blow up two points on the 
divisors $D_s$ and $D_u$ to get ${\cal M}_2$ as shown in Fig.~3c).  This gives 
us two of the 
divisors with self-intersection -2 necessary for ${\cal N}=2$ non-critical 
strings.  We produce the other two such divisors by locating these blow-ups 
pairwise on $D_v$ and $D_u$.  As can be seen from fig.~3a), this last
operation  
corresponds to placing two pairs of M-theory 5-branes at identical $K3$ 
positions.  Figs. 3b) and 3d) illustrate how the construction looks almost 
identical starting from the dual $(12,12)$ model with coupling $\l_2 = a_d$.

\lref\strominger{A. Strominger, {\it Open P-Branes}, hep-th/9512059,
\plb{383} (1996) 44.}

It is interesting to contrast the origins of the exceptional divisors in
the two electric-magnetic dual $(10,10)$ models of figs. 3a) and 3b).  For 
fig.~3a) shrinking the exceptional divisors $D_s$ and $D_t$ corresponds to 
strong coupling singularities inside each end-of-the-world nine-brane, and
shrinking the exceptional divisors $D_v$ and $D_u$ corresponds to ${\cal N}=2$
non-critical strings appearing from 
overlapping 5-branes as in ref.~\strominger.  Fig.~3b) gives us the 
complimentary picture where these two seperate phenomena are exchanged! 

	To continue our quest to link ${\cal M}_2$, the F-theory dual
of the electric and magnetic dual $(10,10)$ models, with $\overline
{\cal M}_2$, the F-theory  
description of the $T^4/\ZZ4$ orientifold, we need only stare at Figs.~3c) and
4c) (or alternatively 3d) and 4d)).  Clearly, blowing down the exceptional
divisors $D_{u,v,s,t}$ to $A_1$ singularities does the job. Let
us examine how the relevant couplings behave. As we mentioned earlier,
blowing up the base $F_0$ of ${\cal M}_1$, will change the K\"ahler class of 
the base.  The new K\"ahler class ($D_i$ are the exceptional divisors
from our blow-ups) is
\eqn\kclass{
K = a_h\left(D_s + D_1 + D_4\right) + a_d\left(D_u + D_1 + D_3\right)
-\sum_{i=1}^4 \phi_i\left(D_i\right),
} 
where $D_i$ are the exceptional divisors from our blow-ups.  This gives
the areas of the divisors of interest as
\eqn\areas{
\twoeqsalign{
{\rm area}\left(D_s\right) &= a_d - \phi_1 - \phi_4,\qquad&
{\rm area}\left(D_t\right) &= a_d - \phi_2 - \phi_3, \cr
{\rm area}\left(D_u\right) &= a_h - \phi_1 - \phi_3,\qquad&
{\rm area}\left(D_v\right) &= a_d - \phi_2 - \phi_4,\quad
{\rm area}\left(D_i\right) = \phi_i.
}}
and a volume for the base proportional to
\eqn\vol{
2 a_h a_d - \sum_{i=1}^4 \phi_i^2.
}
Requiring that all the areas be positive implies that $a_h$ and $a_d$
are bounded from below by the vevs $\phi_i$.  Thus, as we asserted
before, for either 
of the dual $(10,10)$ models to be weakly coupled requires the $\phi_i$s to
be small (the couplings $\l_{1,2}$ are still proportional to $a_{h,d}$), which
will certainly not be the case if we take the limit of ${\cal M}_2$ which
matches $\overline {\cal M}_2$.

\topinsert{
\centerline{\epsfxsize=4.5in
\epsfbox{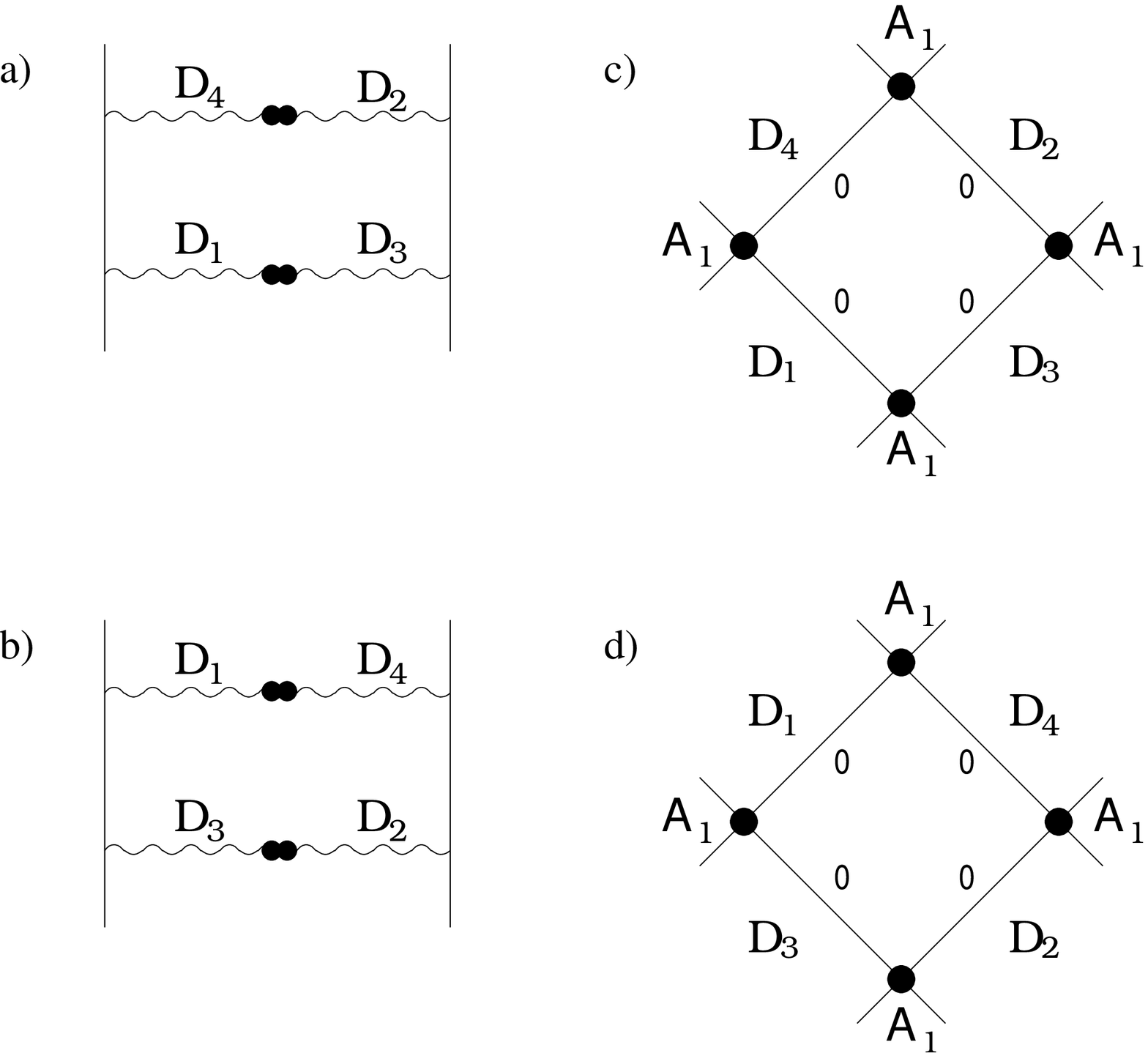}}}
{\baselineskip 10pt
\noindent{{\bf Figure  4.} \sl 

a) Same as fig. 3a), except that we have placed the M-theory 
5-branes on top of each other and gone to a strong coupling
point on both 9-branes.

b) Same as fig. 3b),  except that we have placed the M-theory 
5-branes on top of each other and gone to a strong coupling
point on both 9-branes (not the same 5-branes as in fig.~4a).

c) Same base as in fig. 3c), were have blown down all the -2 divisors
to produce $A_1$ singularities.  Note that the blown-down $D_u$ adds 
$1 \over 2$ to the self-intersection numbers of $D_{1,2}$ and
similarly for the other -2 divisors.

d) Same base as fig. 4c). This base describes F-theory duals to both
figs. 4a) and  
4b).}}\endinsert 

\lref\fontetal{G. Aldazabal, A. Font, L.E. Ibanez, A.M.Uranga,
G. Violero, {\it Non-Perturbative Heterotic D=6,4 Orbifold Vacua},
hep-th/9706158.}

	This leaves us with a puzzle.  In section~2, we described a formulation
of the $T^4/\ZZ_4$ orientifold with 5-branes and 9-branes.  In this
formulation the bulk fields describe a Type I theory for which it is 
possible to have 
the string coupling everywhere strong.  This implies~\edjoe\ a weakly coupled 
dual $Spin(32)/\ZZ_2$ heterotic theory (for an actual construction of this
dual, see ref.~\fontetal).  We just showed, however, that
the two heterotic dual D-strings one can construct are both strongly coupled
in the orientifold limit.  The answer to this puzzle lies in the final step
of our demonstration.
  
\subsec{A new heterotic string theory}

For the last step in our demonstration, we will now show how to relate the
coordinates used in section~4 to describe F-theory on $\overline {\cal M}_2$, 
with the coordinates for ${\cal M}_2$ which connect naturally with the
$(10,10)$ model. From fig.~4c),d) we see that in terms of the divisors 
$D_i,\, i=1,\ldots,4$, $\overline {\cal M}_2$
looks very much like an F-theory which could have a weakly coupled
dual heterotic theory. (Recall that although the four $A_1$
singularities are there the physics from the perspective of the full
F-theory is non-singular.) We propose a different heterotic string
theory in which 
the (dual) strings are obtained in much the same way as that in which
$het_{1,2}$ were obtained to describe M-theory on $K3\times S^1/\ZZ_2$.

 	Let us observe that for F-theory on $\overline {\cal M}_2$, as shown
in fig.~4c)d), the divisors $D_i$ all have self-intersection number 0
(this number 
was raised from -1 when the $A_1$s were blown down).  Also, the presence
of the $A_1$ singularities signals that in this limit, $D_1 \cdot D_3 = 
{1 \over 2}$.  It is
natural to suspect that the divisors $2 D_i$ (the extra factor
of two guarantees integer intersection numbers) 
can be associated with two new divisor classes
with areas $\tilde a_h$ and $\tilde a_d$ such that in the limit where
the exceptional divisors $D_{u,v,s,t}$ shrink to zero size we
have~\foot{The factors of two are really introduced because we are interested
in the generic divisor class. Recall that in
the orbifold $\IP^1(z)\times\IP^1(w)/((z,w)\to (-z,-w))$, the divisors
$z=\hat z\neq 0,\infty$ 
and $w=\hat w\neq 0,\infty$ intersect twice, while $[z=0]\cdot[w=0]=1/2$.}
\eqn\areadivnew{
\tilde a_d = {\rm area}(2D_3) = {\rm area} (2D_4)\,,\quad
\tilde a_h = {\rm area}(2D_1) = {\rm area} (2D_2)\,.
}  
These new divisor classes would then provide us, upon wrapping
D3-branes on them, with two new D-strings ${\tilde {het_{1,2}}}$ with couplings
as in eq.~\couplingdiv\
\eqn\couplingdivnew{
\tilde\l_1 \propto \tilde a_h\,,\quad \tilde\l_2 \propto \tilde a_d.
}

To make this more precise, we can write down these two new 
divisor classes by performing a change of basis on the K\"ahler moduli 
space of ${\cal M}_2$.  The new divisor classes can be represented
as
\eqn\newdivclass{
\eqalign{
\big[ \tilde {D_h}\big] &= \big[{ 2 D_1 + D_s +  D_u} 
\big] = \big[ 2 D_2 + D_t +  D_v\big];\cr
\big[\tilde D_d\big] &= \big[ 2 D_3 + D_t + D_u \big] =
\big[2 D_4 + D_s + D_v \big]}
}
and the K\"ahler class can be rewritten as
\eqn\newclass{
K = {1\over 2}\left(
{\tilde a}_h (\tilde D_d) + 
{\tilde a}_d (\tilde D_h) -
a_s(D_s) - a_t(D_t) -
a_u(D_u) - a_v(D_v)\right).  
}
Taking the limit ${\cal M}_2\to \overline {\cal M}_2$, a quick
computation shows that the volume of the base for $\overline {\cal M}_2$ is 
proportional to 
\eqn\newvol{
{\tilde a}_h{\tilde a}_d.
}
Thus, in F-theory on $\overline {\cal M}_2$ we have a vacuum with a 
base very similar to that of the F-theory dual for the $(12,12)$
model.

The conjectured dual heterotic theory on $\tilde K3$ 
would be that of a theory which, in terms of M-theory on $\tilde
K3\times \tilde S^1/\ZZ_2$, has a heterotic string $\tilde het_1$
obtained by wrapping the membrane around $\tilde S^1/\ZZ_2$. A second
heterotic string, $\tilde het_2$, 
is obtained by wrapping the 5-brane on $\tilde
K3$. Finally, because of the electric-magnetic duality in M-theory
between membranes and 5-branes, the two heterotic strings are dual.
The two couplings are given in terms of the volume and radius, in M-theory
units, of the $\tilde
K3$ and $\tilde S^1/\ZZ_2$ respectively, as
\eqn\elmagnnew{
\tilde \l^2_1=(\tilde \l^2_2)^{-1}=\tilde R/\tilde V\,.
}

There is of course a crucial difference between $\overline {\cal M}_2$ and
${\cal M}_1$, the orbifolding!  This is best understood by looking at
the  change 
of coordinates implied in eq.\newclass.  The exceptional divisor 
 $D_1 + {1\over 2} D_s + {1\over 2}D_u$ is the divisor which
looks like $D_1$ when we shrink $D_{s,t,u,v}$, but it is half a member
of the divisor class $\left[\tilde D_h \right]$.  This means that our new
coordinates are double valued.
Thus the single valued domain when the divisors
$D_{s,t,u,v}$ are small is really $\left(\IP^1\times\IP^1\right)/\ZZ_2$ with 
slightly blown up $A_1$ singularities, exactly our description for
$\overline {\cal M}_2$.  Also, notice that the volume we
compute for $\overline {\cal M}_2$ in eq.\newvol\ is exactly half the volume
one would get using the K\"ahler class, in section 5.1, for the
$(12,12)$  model.

\subsec{The new heterotic string in the $(11,11)$ model}

	It is interesting to note that we can understand the existence 
of the new heterotic strings, $\tilde het_{1,2}$, in terms of the strongly
coupled heterotic $E_8\times E_8$ theory compactified on $K3$ with
instanton embedding $(11,11)$~\foot{This model was studied in detail
in~\louis.}. (For more details,
see~\ref\bgm{P. Berglund, E. Gimon and D. Morrison, work in progress.}.)
The dual F-theory vacuum is that of an
elliptic Calabi--Yau with base $F_0$ blown up at two points.  It can be reached
from ${\cal M}_2$ in two ways by either blowing down $D_1$ and $D_2$, or
by blowing down $D_3$ and $D_4$.

	It is sufficient to study the scenario where $D_1$ and $D_2$ are
blown down.  At this point, all the remaining exceptional divisors,
$D_{s,t,u,v,3,4}$ have self-intersection number -1.  There are now three
different ways to shrink two divisors and obtain a model with a base $F_0$;
i) shrink $D_3$ and $D_4$, ii) shrink $D_u$ and $D_v$ and finally,
iii) shrink $D_s$ and $D_t$.  In terms of the $(11,11)$ heterotic model, these
represent three seperate methods to recover the $(12,12)$ model with its two
dual strings.  Each method preserves two out of {\sl three} dual 
strings.  These strings are associated with the divisor classes:
\eqn\threeclass{
\[D_s + D_4\],\quad \[D_u + D_3\],\quad {\rm and} \quad \[D_s + D_u\]
}
The first two strings are familiar to us as descendants of $het_{1,2}$ in the
$(10,10)$ model. They represent the heterotic string and wrapped 5-brane
on the $(11,11)$ background.  The third string is an entirely new
object, whose
heterotic origins should correspond to a bound state of the last two.   After
blowing up the points to get $D_1$ and $D_2$ it will correspond to
$\tilde {het_1}$. 
Similarly, we can see $\tilde {het_2}$ in the $(11,11)$ model, reached
by blowing down the exceptional divisors $D_{3,4}$.

\newsec{Conclusions}

\lref\romans{L.J. Romans, \npb{276} (1986) 71.}

We have shown that the F-theory description of a IIB $T^4/\ZZ_4$ orientifold is
given in terms of an F-theory orbifold,  $\overline {\cal M}_2={\cal
M}_1/\ZZ_2$  
where ${\cal M}_1$ is the Calabi--Yau vacuum used to describe the F-theory
corresponding to the GP $\ZZ_2$ orientifold.  The appearance of an 
F-theory orbifold in the process has interesting implications, beyond the 
immediate scope of the specific models involved.

	If we look at the volume formula, eq.\vol, for the base of the 
Calabi-Yau ${\cal M}_2$, we see that for fixed volume, the tensor scalars  
describing the K\"ahler moduli space are constrained to sit on a hyperboloid.  
This is entirely consistent with the $SO(1,n_T)$ (here $n_T = 5$)
structure of the  
six-dimensional supergravity tensor scalar moduli space described in 
refs.~\refs{\romans,\sagnotti,\ruben}.  We recover a heterotic description of
the six-dimensional theory when move very far out along one of the branches
of the hyperboloid.  This corresponds to shrinking the divisors $D_i$, with
self-intersection -1, to recover the $(12,12)$ heterotic model.  What we have
discovered is another type of limiting process, different from the one we
just described, which will also recover a heterotic description.

	We found that when we shrink two-cycles with self-intersection -2, a 
process akin to shrinking two-cycles of self-intersection -1 can
happen.  We can rewrite our 
F-theory Calabi-Yau vacuum as the orbifold (with action of order 2) of another 
Calabi-Yau model
whose base contains shrinking two-cycles with self-intersection -1. 
So not only can we recover a perturbative heterotic picture when we
move far out 
along the branches of the hyperboloid constraining the scalars of the 
${\cal N}=1$ tensor multiplets. But we can also recover a perturbative
heterotic 
picture when we are near some of the points in the interior of the hyperboloid,
where two-cycles of self-intersection -2 shrink down.  The implication is that
models were two-cycles with self-intersection $-n$ ($n > 2$) shrink down,
should also 
have a well-defined perturbative heterotic description.  The key would be to
rewrite the corresponding F-theory Calabi-Yau as the orbifold, with
elements of 
order $n$, of another Calabi-Yau with shrinking two-cycles of
self-intersection -1  
and then to connect this later Calabi-Yau, via orientifolds, to a perturbative
heterotic description.  This is a fine illustration of the complementarity
of F-theory, orientifold, and heterotic vacua in string theory.

\vskip5mm
\noindent{\bf Note Added:}
After this work was completed there appeared a paper which studies
the type IIB orientifold $T^4/\ZZ_4$~\ref\tye{Z. Kakushadze, G. Shiu,
S.-H. H. Tye, {\it  Type IIB Orientifolds with NS-NS Antisymmetric
Tensor Backgrounds}, hep-th/9803141.}.

\vskip5mm
\noindent{\bf Acknowledgments:}
We thank A. Hanany, J. Polchinski, A. Sen, R. Minasian, M. Gremm and
in particular C.~Johnson, S. Katz and D. Morrison for useful
discussions.  
P.B. would like to acknowledge the hospitality of the Mittag-Leffler
Institute, Stockholm, Uppsala University, and the
University of California, Berkeley where some of this work was carried
out.
The work of P.B. was supported in part by the National
Science Foundation grant NSF  PHY94-07194.
The work of E.G. was supported in part by the Department of Energy
grant DE-FG03-92-ER40701.
\vfill\eject

\appendix{A}{}

In this appendix we present a more detailed analysis of the
deformations of F-theory on $\overline {\cal M}_2$ corresponding to
the blow-ups of the fixed points of the $\ZZ_4$ orientifold. We also
study gauge enhancement for F-theory on $\overline {\cal M}_2$.

\subsec{Blowing up the orientifold points}

The six $\ZZ_2$ fixed points in the orientifold picture 
correspond to points in the base at
\eqn\ztwo{
 (\tilde z,\tilde w),(\tilde z,-\tilde w),(0,\tilde w),
(\infty,\tilde w),(\tilde z,0),(\tilde z,\infty), 
}
while, to match with the $A_1$ singularities, the four $\ZZ_4$ fixed points 
are at,
\eqn\zfour{
 (0,0),(\infty,0),(0,\infty),(\infty,\infty).
}
We know, from ref.~\senstuff, that these fixed points must be associated
with the crossings of the zeroes of $h(z,w)$.  We can match with the fixed
points above by choosing $Q(z,w) = zw$ in eq.~\hetazfour.  This gives
us the initial form for $h$ where the orientifold points are blown down:
\eqn\hinit{
h_0 = zw(z^2 - \tilde z^2)(w^2 - \tilde w^2)
}

Most of the blow-up modes in the $\ZZ_4$ orientifold can now be encoded 
via deformations of $h_0(z,w)$, in analogy
with the situation for the $\ZZ_2$ orientifold. The analysis for the
$\ZZ_2$ points is exactly the same as in the $\ZZ_2$ orientifold.
For the points $(\tilde z,\tilde w)$ and $(\tilde z,-\tilde w)$ we
have the following deformations 
\eqn\ztwodefa{ 
\eqalign{
\d h=
z w ( & {\a_{11}\over 2} ((z-\tilde z)(w-\tilde w) + 
(z+\tilde z)(w+\tilde w)))  \cr 
& + {\a_{12}\over 2} 
((z-\tilde z)(w+\tilde w) + (z+\tilde z)(w-\tilde w))). \cr}
}
The deformations of $h_0$ associated to the points $(0,\tilde w)$,
$(\infty,\tilde w)$ and 
$(\tilde z,0)$, $(\tilde z,\infty)$, respectively, are given by
\eqn\ztwodefb{
\eqalign{\d h =
&w (z^2 - \tilde z^2) ( {\b_{11} \over 2}
((w-\tilde w) + (w+\tilde w)) + 
{\g_{11} \over 2} (z^2(w-\tilde w) + z^2 (w+\tilde w))) \cr
\d h=
&z (w^2 - \tilde w^2) ( {\b_{22} \over 2}
((z-\tilde z) + (z+\tilde z))+
+ {\g_{22} \over 2} (w^2(z-\tilde z) + w^2 (z+\tilde z))). \cr}
}

The $\ZZ_2$ twisted sector blow-up modes of the the orientifold $\ZZ_4$ points
(those not linked with any tensors) are inherited directly from the
structure of ${\cal M}_1$.   These deformations for the crossing zeroes of 
$h_0$ at $(0,0)$, $(0,\infty)$,
$(\infty,0)$, and $(\infty,\infty)$ are, respectively. 
\eqn\zfourdef{
\eqalign{
h=
&(z^2 - \tilde z^2) (w^2 - \tilde w^2) ( z w + \d_{11} 1
 + \d_{12} w^2 + \d_{21} z^2 + \d_{22} w^2 z^2)}
}

As discussed, these four points also have non-toric deformations which come
in pairs with the ${\cal N}=1$ tensor multiplets.  To realize these
deformations 
we make a change of coordinates given by
\eqn\transform{
a\cdot b = c^2\,,\quad {\rm where} \quad
a = z^2,\quad
b = w^2, \quad
c = zw~.
}
The second set of blow-up modes for the $\ZZ_4$ orientifold points is
then given by the deformation of the four $A_1$ singularities as follows
\eqn\singquotdef{
(a -\l_{12})\cdot (b - \l_{21}) + \l_{12} \l_{21} - \l_{22}^2(a \cdot b) 
= (c-\l_{11})(c+\l_{11})~.
}

\subsec{Gauge Enhancements}

Given the above analysis, we now want to describe the gauge
enhancement occuring  
when a collection of 7-branes are aligned with an $O7_4$ plane using F-theory.
We will illustrate how to get the various enhancement patterns related to those
in fig. 1 (which only describes enhancements in terms of the 5-9 picture).  Our
starting point is situation A, where the collection of 7-branes is in the
bulk and carries a $USp(2n)$ gauge group. 
As in section 4.3,  we can collect several 7-branes on top of each
other by setting $z_1 = \ldots = z_n$ in the defining equation for
$\overline {\cal M}_2$~\hetazfour.  If we use our new coordinates
$(a,b,c)$, and define $a_i = z_i^2, \tilde a = \tilde z^2, b_i = w_i^2,
\tilde b = \tilde w^2$, this collection of 7-branes yields a discriminant
of the form
\eqn\discr{
\Delta \sim (a - a_1)^{2n}
} 
which generically is non-split~\bershad, giving us the requisite
$USp(2n)$ gauge group.

	Next, we would like the collection of 7-branes to approach an
$O7_4$ plane.  If we rewrite $h_0$ as 
\eqn\rewrite{
h_0 = c(a - \tilde a)(b - \tilde b)
} 
then the $O7_4$ planes are located near $a = 0,\infty$ and $b = 0,\infty$.
For the purposes of this discussion, we will consider the $O7_4$ 
plane near the $a = 0$.  To be able to study this case in the most detail, we 
have to deform the base of $\overline {\cal M}_2$~\foot{Recall that the
complex structure deformation of the $A_1$ singularity accounts for
two of the four scalars in the corresponding hypermultiplet. The
remaining two are R-R and NS-NS two-form fluxes. We 
argued that in the 7--7' 
picture of the $\ZZ_4$ orientifold, the NS-NS scalar is
non-zero due to a non-vanishing 
two-form flux, while the remaining three scalars are zero. Thus, we do
not have a direct correspondence between this non-zero scalar and that
of a complex structure deformation of the $A_1$. However, we will
assume that as far as the gauge symmetry is concerned the effect is
the same, as long as we avoid complicated singularities which would
give rise to tensionless strings.}.  The relevant deformations of $h_0$ 
for $a=0$ can be read off from~\ztwodefb\ and \zfourdef\ .  We define the
deformed $h$ as
\eqn\hdef{
h_d =
(a-\tilde a)\{ \d_{12} b^2 + b(c+\b_{11} + \d_{11} - \d_{12}\tilde b)
- \tilde b(c+ \d_{11}) \},
}
For $a=0$ there are two $A_1$ singularities in the base located at
$b=0$ and $b=\infty$, respectively. To deform only with respect to these
singularities we restrict eq.~\singquotdef\ such that
\eqn\adeform{
\l_{11} = \l,\quad 
\l_{12} = \l',\quad
\l_{21} = \l_{22} = 0~.
}

We can now consider what happens when the collection of 7-branes
approaches the deformed $O7_4$ described above.  We do this
by letting $\a_1\to \l'$.   Inserting this in  our eqn.~\deltasutwob\ for the
discriminant, $\Delta$, and using eqns. \singquotdef\ , \adeform\ and \hdef\
we have 
\eqn\deltazfourb{
\Delta =
b^{-2n} (c-\l)^{2n}(c+\l)^{2n}{\tilde\eta}^2(a,b)
\{-9 h_d^2(a,b,c) + {\cal O}(c-\l)(c+\l)\}~,
}
where $\tilde\eta(a,b)=C\prod_{i=n+1}^4
(a-a_i) \prod_{j=1}^4 (b-b_j)$.
Thus, by tuning one parameter ($a_1 \to \l'$), the discriminant takes the form
\eqn\deltazfourB{
\Delta \sim ({(c-\l)(c+\l)\over b})^{2n}
}
The divisor $a = a_1$ has split into two separate divisors, and we
get an enhancement to an $A_{2n-1}\times A_{2n-1}$ singularity locus. 
For generic $h_d$ the gauge group is
$USp(2n)\times USp(2n)$. This corresponds to configuration~B in
fig.~1.  We get matter transforming as $({\bf 2n,2n})$; after higgsing 
$Sp(n)\times Sp(n)$ to
$Sp(n)$ using $({\bf 2n,2n})$,  the  remaining matter is in the $({\bf
 n(2n-1)-1})$ + ${\bf 1}$  representation of $Sp(n)$.

We can choose our deformations such that $h$ becomes a perfect
square,  for either of $c=\pm \l$ or for both. This corresponds to
either (or both) of the $A_{2n-1}$ singularities to be
split~\bershad. Let us rewrite $h_d$ as
\eqn\hsquareC{
h_d = (a-\tilde a)((c-\m)(b + \n)^2 - 
                (c+\tilde \m)(b + \tilde \n )^2)
}
where $\n,\tilde \n$ and $\m,\tilde \m$ are functions of the $\b$'s
and $\d$'s in eq.~\hdef. By setting either $\m=\l$
or $\tilde \m = -\l$, we get a perfect square for
$c=-\l$ or for $c=\l$. In either of those cases, $h$ is
a perfect square and we get an enhancement to
$SU(2n)\times Sp(n)$. The only matter consistent with this enhancement
is $({\bf 2n,2n})+ ({\bf n(2n-1),1})$. This corresponds to configurations 
C and C' in fig~1.  If we let  $\m=\l$ {\it and} $\tilde \m = -l$, then $h$ is 
a perfect square for {\it both} $c=-\l$ and for $c=\l$.  We get a further
enhancement to $SU(2n)\times SU(2n)$. The matter is in the
$({\bf 2n,2n})+ ({\bf n(2n-1),1}) + ({\bf 1, n(2n-1)})$ representations.
This is in agreement with the situation in configuration~E in fig~1. 

Finally, if we relax the condition $a_1 = \l'$ there is still a possibility
for enhancement.  When $a=a_1\neq \l'$ we can use eq.\singquotdef\ to
write $b$ as a quadratic expression in $c$.  Then we can rewrite $h_d$ 
as a quartic in $c$.  Tuning two parameters will make $h_d(c)$ a
perfect square, 
and the $A_{2n-1}$ singularity in eq.\discr\ will now be split.  Thus,
exactly as 
in configuration~D in fig~1, tuning two parameters gives us an $SU(2n)$ gauge
group.

	We have found a complete correspondence between the various gauge
enhancement patterns in the $\ZZ_4$ orientifold on one hand and in
F-theory on $\overline{\cal M}_2$ on the other.  Note that if we keep $h = h_0$
and set $a_1 = 0$,  then our F-theory model will be even more
singular.  It will 
have both ${\cal N}=1$ and ${\cal N}=2$ tensionless strings!  This
does not match with
the $\ZZ_4$ orientifold.  The reason is that
in F-theory we have set all relevant scalars to zero, but in the orientifold 
some scalars (B-field fluxes) are non-zero.  This is not apparent with the set
of variables we are using to describe F-theory, which only describe two out of
the four scalars in any hypermultiplet and hence miss these scalars.
Fortunately, since we 
keep the first two scalars non-zero for the relevant hypermultiplets, our 
analysis is not compromised (a similar issue arises already in
ref.~\senstuff).   

\listrefs
\vfill\eject
\bye